%% file: Thesis.tex
\documentclass{report}

\usepackage[a4paper, margin=2.5cm]{geometry}

\usepackage{parskip} 
\usepackage{setspace} 
\usepackage{emptypage} 
\usepackage{multicol} 
\usepackage{xurl} 

\usepackage{titlesec}
\titlespacing{\section}{0pt}{3.3ex}{2ex}
\titlespacing{\subsection}{0pt}{3.3ex}{1.65ex}
\titlespacing{\subsubsection}{0pt}{3.3ex}{1ex}
\usepackage{color}

\usepackage[english]{babel} 
\usepackage[utf8]{inputenc} 
\usepackage[T1]{fontenc} 
\usepackage[11pt]{moresize} 

\usepackage{graphicx}
\usepackage{transparent} 
\usepackage{eso-pic} 
\usepackage{subfig} 
\usepackage{tikz} 
\usetikzlibrary{}
\graphicspath{{./Images/}} 
\usepackage{caption} 
\usepackage{xcolor} 
\usepackage{amsthm,thmtools,xcolor} 
\usepackage{float}

\usepackage{tabularx}
\usepackage{longtable} 
\usepackage{colortbl}

\usepackage[colorlinks=true,linkcolor=black,anchorcolor=black,citecolor=black,filecolor=black,menucolor=black,runcolor=black,urlcolor=black]{hyperref} 
\usepackage{cleveref}
\usepackage[square, numbers]{natbib} 
\usepackage{fancyhdr} 
\begin{document}

\fancypagestyle{plain}{%
\fancyhf{} 
\fancyhead[RO,RE]{\thepage} 
\renewcommand{\headrulewidth}{0pt}
\renewcommand{\footrulewidth}{0pt}}


\begin{titlepage}

\centering

{\LARGE\textbf{Emotion-Aware Conversational\\[0.2cm]
Recommender Systems: a Case Study}}\\[1.2cm]

{\large \textbf{Author:} Maria Stella Albarelli}\\[6pt]
{\large \textbf{Study Programme:} Computer Science Engineering}\\[6pt]

\vspace{1cm}  
{\Large \textbf{Abstract}}
\vspace{1cm}  
\begin{flushleft}


In recent years, especially during the COVID-19 period, online shopping has seen rapid growth, with users increasingly purchasing items through online platforms. Despite this, the online shopping experience still lacks key elements present in physical stores, such as the opportunity to receive empathic support and dedicated advice from a professional sales assistant. 

This study investigates how an empathic Conversational Agent (CA) can transform the online shopping experience by responding to user emotions with empathy and appropriateness, creating a more natural and humanized interaction. The research focuses on developing Gala, an emotion-aware virtual assistant designed to recommend products from the Galeries Lafayette website. Gala is equipped to recognize users' emotional states through their voice messages, allowing it to respond empathetically basing on perceived emotions. The work started with a set of semi-structured interviews to analyze user needs and define the core functionalities that informed the design of Gala UX and capability. 
Its implementation used the OpenAI API and the Galeries Lafayette API. The recommendation approach follows a Content-Based methodology. Using Natural Language Processing (NLP), the assistant interprets the user's requests and searches items in the product catalogue that align with the specified attributes, such as name, price, and brand. These features generate a smooth natural dialogue and provide product recommendations. 
Subsequently, two phases of user testing were conducted: an initial usability test to evaluate the system usability, and a second user test to compare a standard CA with Gala's emotion-aware version. 

In conclusion, the results highlight the potential of emotion-aware CAs to enhance online shopping by making product selection faster and more engaging. This provides a guided experience similar to that in a physical store.
\\
\vspace{1cm}
\end{flushleft}
\end{titlepage}


\thispagestyle{empty}
\tableofcontents 
\thispagestyle{empty}
\cleardoublepage

%
%
%

\addtocontents{toc}{\vspace{2em}} 

\input{Intro}

\input{literature_review}

\input{design}

\input{implementation}

\input{empirical_studies}

\input{conclusions_future_works}


\addtocontents{toc}{\vspace{2em}} 
\bibliography{Thesis_bibliography} 


\cleardoublepage
\addtocontents{toc}{\vspace{2em}} 

\end{document}

%% file: Intro.tex
\chapter{Introduction}
\label{sec:intro}

A Conversational Recommender System (CRS) is a software that supports users providing personalized recommendations through a multi-turn dialogue. One key feature of CRSs is their ability to provide recommendations targeted to specific tasks. In addition, they play a crucial role in assisting users throughout the decision-making process.

\section{Problem and Contribution}

This project was carried out in collaboration with the French department store Galeries Lafayette \footnote{https://www.galerieslafayette.com/} in Paris, as a part of the internship program I undertook. The general challenge Galeries Lafayette wanted to address was to improve the shopping experience of their customer, introducing innovative and engaging elements, refining what shopping at their stores could feel like. 

The solution aimed to address this problem by designing and implementing an intuitive conversational AI model that could be easy to understand and use. Additionally, the solution must support various forms of interaction, such as voice messages and image sharing, to ensure a smooth conversation and allow the user to interact in multiple ways, receiving accurate responses. 

Each week, was defined a list of goals and features to introduce in the project, to reach the final design. At the beginning of the study, various technologies were explored for integration into the assistant's features, including the use of stable diffusion to apply catalogue products to different categories of models. Virtual try-on was also analyzed to determine whether users could appreciate the ability to virtually try on clothes online, allowing them to assess the fit and adaptability to their body. Unfortunately, these features were not included in the final result due to time constraints and limited knowledge of the technologies. 

The final goal was to create an experience where the user can communicate with an assistant that makes online shopping feel as close as possible to the in-store experience.

The first idea was to create an avatar to assist people during their experience in the store, using Augmented Reality (AR) to introduce new engaging ways to interact with the avatar. The purpose of the avatar was to guide and give information to users inside the store.
In the end, I opted to implement a CRS for online shopping due to limited resources and time constraints. I evaluated that a CRS could be more versatile and easier to test with real users. 



\section{Research Question}

Given the aforementioned scope, I focused on a specific research direction: exploring emotion recognition in CRSs to provide context-sensitive recommendations, aiming to create an online shopping experience similar to the in-store one, with interactions resembling those with real human assistants.

Indeed, due to limited studies on empathic recommender agents within the fashion retail sector, I decided to focus on adapting the conversation flow and the assistant's behaviour in response to the user's emotional state. Analyzing the tone of voice in users' voice messages enables the assistant to interpret emotional nuances, adapting its conversational approach to align more closely with the users' current emotional states. 

Studies highlight that empathic behaviours in AI-driven interactions can foster perceptions of trustworthiness, which is essential for building a relationship between customers and shopping assistants. Trust, in turn, can enhance user experience and influence decision-making processes, potentially leading to greater user satisfaction and increased likelihood of purchase.

Furthermore, relevant research report that emotions can drive purchasing decisions. This interplay between emotion and decision-making underscores the importance of designing Recommender Systems (RSs) that are not only able to suggest products, but also to engage with users empathetically. This leads to the formulation of the following research question:

\begin{center}
\textit{How does the wording of recommendations change when an emotional component is present? How does this alter the user's online shopping experience?}
\end{center}

%% file: literature_review.tex
\chapter{Literature Review}
\label{sec:rel-work}

This section delves into the literature review conducted touching on numerous topics that form the foundation of an emotion-aware conversational shopping assistant. The first part focuses on the keys technologies, including Conversational Recommender Systems (CRSs), Conversational Agents (CAs) and Natural Language Processing (NLP).
The second part of the research explores the concept of 
Affective Computing and automatic emotion recognition.

\section{CRS technologies}
In the context of digital commerce, CRSs are transforming the online shopping experience by enabling personalized and assisted interactions. The ability of these systems to integrate recommendations with human-like conversation offers an alternative to traditional recommendation tools, creating an experience similar to one with an in-store assistant. 

CRSs combine advanced recommendation algorithms, NLP and CAs' features to provide context-sensitive suggestions that respond to specific user needs, enhancing the shopping experience.

\subsection{Recommender Systems}
\label{sec:RS}
A Recommender System (RS) filters and analyzes input data to provide users with hints and suggestions about items that can meet their interests \cite{hutchison_looking_2011}.
Different types of input data are required for RSs to generate recommendations, such as \textbf{Items Data} that is a list of available items, which is the primary input for any recommender algorithm. \textbf{Users Data} which is a list of user attributes, such as gender and age, to tailor recommendations to individual preferences. \textbf{Interaction Data} which includes insights into user opinions on items through their interactions with the system. Finally, \textbf{Context Data} that is a list of attributes related to the context of interactions, determining the appropriate area of interest for recommendations. Examples of contextual attributes are geographical area and day of the week.

Recommender algorithms are, in turn, classified into two categories: 

    \begin{itemize}
        \item \textbf{Non-personalized recommendations}:
         Provide the same suggestions to all users, such as trending movies or music. 
         \item \textbf{Personalized recommendations}: Offer suited suggestions based on individual user data.
    \end{itemize}

Personalized recommendation techniques can be further categorized, the first is the\\
\textbf{Content-Based Filtering} technique, which provides recommendations based on items that are aligned with user's preferences, requiring a list of quality attributes for each product. For instance, a garment can be characterized by genre, size, category and colour. Another type of personalized recommendation technique is the  \textbf{Collaborative Filtering} which relies on the opinions of a community of users, it recommends what similar customers bought or liked \cite{hutchison_looking_2011}. This latter technique is categorized into:

\vspace{.2cm}

    \begin{itemize}
        \item \textbf{User-Based}:
        Based on users with similar tastes.
        \item \textbf{Item-Based}:
        Based on item similarity according to user opinions.
        \item \textbf{Matrix Factorization and Factorization Machines}:
        Techniques to decompose large user-item matrices into latent factors.
        \end{itemize}

Then there is the \textbf{Context-Aware Recommender Systems (CARS)} technique that extends collaborative filtering by incorporating context to improve the quality of recommendations. Lastly, there are the \textbf{Hybrid Approaches} that merge and enhance the capabilities of content, collaborative, and context-based techniques.

\subsection{Natural Language Processing}
 A CA tries to replicate human conversations through the use of NLP. It is a specific field of artificial intelligence and its goal is to enable computers to understand, interpret, and respond to natural language in meaningful ways. It analyzes large amounts of textual data for applications such as speech recognition, machine translation, sentiment analysis, and text generation \cite{khur_natural_2023}. \\
NLP is classified into two parts:

    \begin{itemize}
        \item \textbf{Natural Language Understanding (NLU)}: Allows machine to understand human language by extracting concepts, emotions and keywords.
         \item \textbf{Natural Language Generation (NLG)}:
         Creates phrases and sentences meaningful  for the context of use. It happens in three phases: identifying the goals, planning on how goals can be achieved and realizing a plan.
    \end{itemize}

\subsubsection{Large Language Models}

With the introduction of Large Language Models (LLMs), NLP capabilities have expanded. LLMs use advanced architectures like Transformers and extensive datasets to enhance NLU and NLG performance. LLMs represent a sophisticated category of AI systems, characterized by their ability to understand, generate and interpret human language with exceptional precision. They are trained on massive text data, such as GPT-3 \cite{brown2020languagemodelsfewshotlearners} and LLAMA \cite{touvron2023llamaopenefficientfoundation}. 

LLMs can process and generate language with greater accuracy and contextual awareness, making them indispensable in modern NLP applications \cite{zhao_survey_2024}. The evolution of LLMs has been marked by significant milestones, like the development of the model Generative Pretrained Transformer (GPT), which has had a lot of improvements in the last years. 

\textbf{GPT-1} \cite{zhao_survey_2024} was the first model in the series to introduce the Transformer architecture, demonstrating how pre-training on raw text data can improve NLP tasks. 

Next, \textbf{GPT-2} \cite{zhao_survey_2024} expanded the parameters to 1.5 billion, enabling coherent text generation on varied topics and showcasing the potential of LLMs for unsupervised learning.

With \textbf{GPT-3} \cite{ZHANG2021831}, equipped with 175 billion parameters, new capabilities emerged, such as in-context learning, allowing the model to perform complex tasks without specific training. This model marked a major leap in application versatility.

Following GPT-3, \textbf{InstructGPT} \cite{zhao_survey_2024} was trained with human feedback using Reinforcement Learning from Human Feedback (RLHF), making it more responsive to human preferences and values. From this model, ChatGPT \cite{casheekar_contemporary_2024} was developed, optimized for conversational interactions, and able to handle multi-turn dialogues with structured, coherent responses.

\textbf{GPT-4} \cite{zhao_survey_2024} introduced multimodal capabilities, with the ability to understand both text and images, further improving on complex reasoning tasks and content safety. 

\textbf{GPT-4 Turbo} \cite{zhao_survey_2024} optimized GPT-4’s performance with extended context and reduced costs, making it ideal for scalable, high-efficiency applications.

Finally, the latest version, \textbf{GPT-4o} \cite{chugh2023evolution}, is notable for its advanced capability to handle and integrate various types of data, including text, images, video, and audio. Furthermore, there is the possibility to incorporate models that are fine-tuned to meet the specialized demands of specific applications and industries.

\subsection{Conversational Agents}
\label{ssec:ca}
CAs are virtual assistants that communicate using human-like language, to create a more natural dialogue with users. These intelligent systems are based on AI features to understand and react to user requests. Also, CAs exploit NLP and Deep Learning technologies to understand human language \cite{casheekar_contemporary_2024}. 



The origin of CAs dates back to the middle of the 20th century, when the first chatbot created was ELIZA, developed in the 1960s by Joseph Weizenbaum. ELIZA was a rule-based system that captured the input, rephrased it, and tried to match keywords with a pre-defined set of responses ~\cite{singh_survey_2022}. 

With the arrival of new technologies, like cloud computing and large-scale dataset, new chatbot platforms were introduced, such as ChatGPT \cite{casheekar_contemporary_2024}, which is the OpenAI AI-powered virtual conversational agent, introduced the first time in the November of 2022. ChatGPT generates text based answers using the GPT neural network architecture. This architecture is formed by multiple layers of self-attention mechanism and learns from a large amount of text data.

Most of all, the latest version developed by OpenAI, ChatGPT-4o \cite{OpenAI2024}, shows significant progress. In fact, this latest version generates increasingly coherent and contextually relevant responses, consequently improving human-computer interaction. GPT-4o was trained using the RLHF method. This is a method where machine learning models are trained using feedback from humans to improve their performance \cite{OpenAI2024}.

\subsection{Conversational Recommender Systems}
A CRS combines recommendation metrics and NLP techniques to provide different types of suggestions, based on the user's needs and preferences. In the world of fashion retail a CRS has the function of recommending relevant products and convincing the customer to buy the product, just like a shopping assistant \cite{pramod_conversational_2022}. It is worth noting that RSs play an important role in the online shopping field; for instance, they drive up to 35\% of Amazon \footnote{https://www.amazon.it/} sales \cite{deldjoo2023reviewmodernfashionrecommender}.

CRSs use the main recommendation techniques: Collaborative Filtering, Content-based, Context-Aware and Hybrid Approaches (Section \ref{sec:RS}). An important feature of CRSs is their capability to create a multi-turn conversational interaction. Unlike the basic digital assistants, which provide one-shot Q\&A-style recommendations, CRSs can respond to recommendation requests, keeping track of the conversation history and the current state.

In fact, the most used CRSs model to gather user preferences is the interactive recommender model, which emphasizes the continuous interaction between the user and the system to improve the quality of recommendations \cite{pramod_conversational_2022}. The interactive model can be:

    \begin{itemize}
        \item \textbf{Utility based}:
        The utility of each item is evaluated using a multi-attribute method, allowing users to express their preferences.
        \item \textbf{Dialog based}: Uses a natural language based conversation in spoken or typed form to collect user utterances and create a user profile to better customize recommendations.
        \item \textbf{Critiquing based}: Gathers users ratings and critiques about a product to provide hence data-driven.
        \item \textbf{Constraint based}: Takes into account user and product constraints to give recommendations that meet those constraints.
        \end{itemize}



The performance of RSs is typically evaluated using metrics such as precision, recall, F-measure, RMSE (Root Mean Squared Error), and MAE (Mean Absolute Error).
Additionally, user-centric evaluation frameworks, like ResQue, assess the quality of user experience by measuring factors such as trust, satisfaction, and perceived usefulness \cite{hutchison_looking_2011}.

\section{Affective Computing}
Emotions are fundamental to human interactions, as they allow us to express our feelings and interpret impulses in our relationships with others. The emotions we experience during a conversation can shape its direction, influencing both our words and decisions \cite{lerner_emotion_2015}.

Emotions can be divided in “primary” or “basic” and “secondary”. The term “primary” emotions refers to emotions which are supposed to be innate. They evolved through phylogeny to allow quick, reactive responses to immediate threats. Instead, “secondary” emotions like “relief” or “hope” are assumed to arise from higher cognitive processes, based on an ability to evaluate preferences over outcomes and expectations. For “secondary” emotions are intended “adult” emotions \cite{prendinger_intelligent_2008}. 

\label{par: Big Six}
In the 20th century, the psychologist Paul Ekman identifies six basic emotions and he suggested that they were experienced in all human cultures. Since 1996, this set of emotions has been known as the “Big Six”, underscoring the significance of his model \cite{Cornelius1997TheSO}. The Big Six are: happiness, sadness, fear, disgust, anger and surprise \cite{ekman_argument_1992}.

    \begin{itemize}
        \item \textbf{Happiness}:
        A pleasant emotion that is characterized by feelings of joy, contentment, gratification, satisfaction and well-being.
        \item \textbf{Sadness}: Considered to be one of the basic human emotions and it is a natural
        response to situations involving psychological, physical or emotional pain or loss of something.
        \item \textbf{Fear}: One of the most basic human emotions that can also play an important role in survival. Fear helps to protect us. It makes us alert to danger and prepares us to deal with it.
        \item \textbf{Disgust}: Can originate from an unpleasant smell, taste or sight. Researchers believe that this emotion evolved as a reaction to foods that might be harmful.
        \item \textbf{Anger}: Can be a powerful emotion characterized by feelings of agitation, hostility and frustration.
         \item \textbf{Surprise}: It is characterized by a physiological startle response following something unexpected. This type of emotion can be positive, negative, or neutral. 
        \end{itemize}

From the concept of Emotion, the concept of Empathy can be derived. Empathy can be defined as 

\begin{quote}
The feeling by which one understands and shares another person’s experiences and emotions \cite{spitale_towards_2020}.
\end{quote}

Empathy plays a fundamental role in the user's experience. The psychologist Baron-Cohen, in particular, distinguishes between cognitive and affective empathy. \textbf{Cognitive empathy} involves understanding how another person feels, whereas \textbf{affective empathy} is an active emotional response to another person’s emotional state.

Emotion-Aware Conversational Recommender Systems can be regarded as a subfield of Affective Computing, a broader discipline defined by Rosalind Picard in her foundational work, Affective Computing (1997) \cite{picard2000affective} as
\begin{quote}
Affective Computing is the study and development of systems and devices that can recognize, interpret, process, and simulate human emotions. 
\end{quote}

To recognize emotions, Emotion-Aware Conversational Recommender Systems employ the process of automatic emotion recognition. This capability allows agents to respond in a proper way, improving interaction quality and fostering a more assisted experience.

\subsection{Automatic Emotion Recognition}

Information about a person’s emotions can be gathered from various cues, such as tone of voice, facial expressions, gestures, and posture. 

Initially, Paul Ekman concentrated specifically on emotions that were expressed by humans through facial expressions \cite{universal_and_cultural_differences_1987}. However, his research was easily expanded to include other communication channels. Subsequently, he investigated the recognition of the Big Six through vocal expressions \cite{sauter_cross-cultural_2010}.

Studies indicate that, according to \cite{chibelushi2003facial}, voice intonation is responsible for about \textbf{85\%} of the message perception in verbal information transmission, while actual words account just for the \textbf{15\%}. For this reason, I chose to focus exclusively on vocal tone, as it provides the ability to express and to understand information not openly communicated as factual content. 

I examined the content of the speech in term of meaning, the prosody of the speech, and the sentiment of the sentences of the speech to understand the affective state of the user.  Voice detection is also a non-intrusive method for real-time emotion detection, which only requires users to send voice messages through the microphone of the device. 

Moreover, recent advancements in machine learning and NLP have led to the development of sophisticated models capable of detecting and interpreting emotional cues from text and speech.

\subsubsection{Speech Emotional Corpora}

Enabling the recognition of specific emotions requires specialized datasets to train the system effectively. Emotional corpora, which are collections of affective materials such as audio recordings, are essential for this purpose.
The quality of an emotional corpus is evident in the communicative effectiveness of its samples, which can significantly influence research outcomes across various fields. Thus, selecting and developing high-quality corpora is essential to avoid drawing incorrect conclusions. 

According to the literature \cite{article, campbell_databases_nodate, gadhe_emotion_2015}, speech emotional corpora are defined by specific characteristics that make them more effective for certain tasks over others.

    \begin{itemize}
        \item {They can include audio recordings with monolingual or multilingual sentences.}
        \item {They can collect different sets of emotions (e.g., the Big Six emotions). }
        \item {They can contain (or not) audio recordings uniformly distributed over emotions.}
        \item {They can include (or not) audio recordings with a set of phrases uniformly verbalized with different emotions.}
        \item {They can be obtained through professional or amateurish recording tools.}
        \item {They can include speech recorded in a fully-setup environment without any noise or in a wild setting.}
        \item {They can contain additional information about the context where speech was recorded, including a description of the situation (e.g., conversational context) or other complementary communication channels (e.g., video).}
        \item {They can collect audio recordings with simulated, induced, or natural emotions.}
        \item {They can contain audio recordings by professional or semi-professional actors or a generic audience with no acting experience.}
    \end{itemize}

Additionally, corpora can include varying numbers of actors with different ages and genders. Most corpora focus on categorical emotions, particularly the Big Six, but different, authors took into account “neutrality” as an supplementary emotional state \cite{inproceedings, interactive_emotional, costantini-etal-2014-emovo, documentation_emotional_speech_danish}. It is also common to find the same sentences expressed in different tones of voice \cite{inproceedings, costantini-etal-2014-emovo, multimodal_emotion-recognition}. This approach aims to base emotion recognition solely on the emotional content of the speech, independent of its lexical elements.

Some of the most famous emotional corpora are: 

    \begin{itemize}
        \item \textbf{DES}: A Danish-language dataset representing anger, joy, neutrality, sadness and surprise \cite{documentation_emotional_speech_danish}.
        \item\textbf{SAVEE}: An English-language dataset representing the Big Six emotions plus neutrality \cite{multimodal_emotion-recognition}.
        \item\textbf{EMO DB}: A German-language dataset representing the Big Six emotions plus neutrality \cite{inproceedings}.
        \item\textbf{EMOVO}: An Italian-language dataset representing the Big Six emotions plus neutrality \cite{costantini-etal-2014-emovo}.
        \item \textbf{Emozionalmente}: An Italian-language dataset capturing the Big Six emotions, along with neutrality. This dataset was developed by Fabio Catania as part of his PhD research at Politecnico di Milano \cite{catania_speech_nodate}. This dataset serves as the emotional corpus used for this project.
    \end{itemize}

\subsection{Automatic Speech Emotion Recognition}

Automatic Speech Emotion Recognition (SER) is an AI technology designed to detect and identify emotions expressed through spoken language. It is commonly approached as a classification task, rooted in the foundational theories of categorical emotion models. By analyzing tone, rhythm, volume, pitch, and other vocal characteristics, SER uses machine learning algorithms and neural networks to infer the speaker's emotional state \cite{SER}. 

The process of SER, shown in figure \ref{fig:brainstorming}, is divided in \textbf{audio pre-processing}, \textbf{audio representation} and \textbf{audio classification}. The initial step, which involves the collection of speech samples, includes various audio cleaning processes, such as noise reduction and normalization, to eliminate unwanted noise from the recording \cite{catania_garzotto}. 

Consider an audio file that includes both the primary sound, such as a voice, and background noise, such as the hum of a fan. The signal spectrum reflects the entire sound, capturing all frequencies and intensities, whereas the noise spectrum isolates just the background noise. To reduce this noise, an average of the noise spectrum is calculated and then subtracted from the signal spectrum. This process suppresses the noise within the signal, ideally leaving the main sound intact. Although this technique is not flawless, it significantly improves clarity by reducing background interference.

During the part of audio representation it is possible to use two types of features: \textbf{prosodic features} and \textbf{spectral features} \cite{SER}.

Prosodic features focus on how words are spoken, considering elements like rhythm, pitch, and pauses, which can help identify basic emotions in speech \cite{catania_speech_nodate}. Spectral features involve transforming the speech signal to analyze its frequency components. An example is Mel-Frequency Cepstral Coefficients (MFCCs), which provide insights into the “power” of vocal sounds over brief time segments. These features are also highly effective in identifying emotions within spoken language \cite{comparative_analysis}.

Once the features have been extracted, machine learning algorithms are employed to classify the emotions. Various models, including Support Vector Machines (SVMs) \cite{valkenborg_support_2023}, Deep Neural Networks (DNNs) \cite{8203938}, and Convolutional Neural Networks (CNNs) \cite{oshea2015introductionconvolutionalneuralnetworks}, are trained on these features to differentiate among emotional states. Each model type offers distinct advantages in terms of accuracy and processing efficiency.

\begin{figure}[h]
    \centering
    \includegraphics[width=15cm]{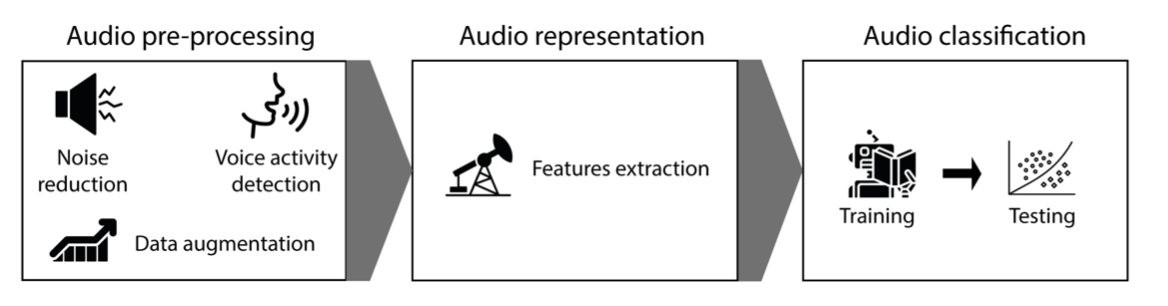}
    \caption{The various steps that constitute a Speech Emotion Recognition system}
    \label{fig:SER}
\end{figure}

Finally, the performance evaluation of a SER system relies on standard classification metrics such as accuracy, precision, recall, and F1-score. Accuracy provides an overall measure of the system's correctness, while precision and recall help to assess the model's ability to make correct predictions and retrieve relevant emotions, with the F1-score balancing these two aspects. Additionally, cross-validation techniques, like k-fold or leave-one-out cross-validation, are used to prevent overfitting, averaging performance across multiple validation sets. Together, these tools provide a comprehensive view of the SER system's validity and reliability \cite{SER}.

\subsection{Adaptive Response}
 Once an emotion is recognized, the virtual agent must determine an appropriate response. It is essential to define \textbf{coping} as the way a person responds to the significance they assign to an event. People are motivated to respond differently based on how they perceive and evaluate the event's importance. For example, events perceived as undesirable but controllable motivate people to create and implement plans to change the situation. In contrast, events viewed as uncontrollable often lead individuals toward avoidance or resignation \cite{multimodal_emotion-recognition, marsella_modeling_2003}.

It is possible to define two different strategies of coping: \textbf{problem-focused coping} and \textbf{emotion-focused coping} \cite{marsella_modeling_2003}.
In the case of problem-focused coping, a person tries to address the cause of the stress or problem directly. The goal is to change the situation to resolve or improve it.
For example, if a test is causing stress, a problem-focused coping strategy might be to study more or take a support class. In fact, if the problem is controllable, problem-focused coping is more useful because it aims to solve the source of the stress.

In the case of emotion-focused coping, instead, the person focuses on managing their emotions rather than tackling the root of the problem, especially if they feel the problem is uncontrollable or unchangeable. For example, if someone misses out on an opportunity, an emotion-focused coping strategy could involve re-evaluating the situation by telling themselves that there will be other opportunities in the future. If the problem is perceived as beyond the person’s control, emotion-focused coping is preferable, as it helps to manage the emotional impact without trying to change external reality.

The literature \cite{paiva_caring_2004} suggests setting specific empathic goals to guide responses when a particular emotion is recognized in the user's behaviour, helping to determine how the system should act accordingly. Some of the goals are:
\label{par: empathetic behavior}

    \begin{itemize}
        \item \textbf{Console}: By making the user feel loved and understood.
        \item\textbf{Encourage}: By providing comments or motivations.
        \item\textbf{Congratulate}: By providing positive feedback on the user's behaviour.
        \item\textbf{Joke}: By doing some humor in order to improve the user's attitude.
         \item\textbf{Calm down}: By providing comments and suggestions to make the user feel more relaxed.
    \end{itemize}

 Furthermore, the paper \cite{marsella_modeling_2003} provides descriptions of several coping strategies that virtual agents might employ:
\begin{itemize}
\label{par: coping strategies}
        \item \textbf{Planning}: This involves creating a plan to overcome or manage the stressor, such as developing a step-by-step solution. This approach is closely tied to goal achievement and effective problem-solving.
        \item\textbf{Positive Reinterpretation}: This approach involves searching for positive aspects or identifying a “silver line” in a stressful situation. It involves reinterpreting the event to emphasize potential benefits or opportunities for growth and learning.
        \item\textbf{Acceptance}: A strategy used when the individual recognizes that a situation is unchangeable. It involves accepting the reality of the event, reducing the emotional impact by removing the pressure to change it.
        \item\textbf{Seeking Social Support}: Engage in others for emotional or instrumental support, which may include desire advice, moral support, or sympathy.
         \item\textbf{Denial/Wishful Thinking}: Avoiding the reality of the situation or believing that things will improve without concrete evidence.
         \item\textbf{Mental Disengagement}: Distracting oneself or detaching mentally from the stressor, often used when avoidance is needed temporarily to cope with overwhelming emotions.

    \end{itemize}

\section{Related Works}

This section provides an overview and analysis of existing Conversational Recommender Systems employed in the fashion world as shopping assistants, but also an emotion-aware conversational agent called Emoty.

\subsection{Chika: a Virtual Agent for e-commerce}
An example of an existing project in the field of chatbots that recommend fashion products is the Virtual Agent (VA) Chika \cite{anastasia_designing_2021} implemented in Shopee's e-commerce platform. This VA is designed to enhance the user experience by addressing common issues such as the cold start problem, data privacy concerns \cite{BELDAD201662, virtual_agents, How_recommendations_affect_impulse_buying}, and lack of social presence in online shopping environments \cite{social_cues}. Chika interacts with users in a conversational manner, helping them to find products, promotions, and similar items based on their preferences \cite{anastasia_designing_2021}.

The VA project employs a User-Centered Design (UCD) methodology combined with a Natural Conversational Framework. The UCD approach ensures that the design process focuses on user needs at every stage, from understanding the context of use to specifying user requirements, creating design solutions, and evaluating the outcomes. The Natural Conversational Framework helps in designing the interaction between the VA and users, making the conversations more natural and seamless \cite{anastasia_designing_2021}.

Despite this, Chika does not allow users to engage in various forms of interaction, such as sending images or exchanging voice messages, relying solely on text messages. This limitation could reduce engagement and the natural flow of conversation.

\subsection{Athena}
Athena \cite{sapna_recommendence_2019} combines a Recommender System with a Fashion-Knowledgeable Component (FKC) into a chatbot. The objective of the project is to provide an real shopping experience through online service. Athena’s RS uses the product inventory of the e-commerce site while its FKC uses fashion information collected from social media, models’ photographs and stylists’ curation of fashion items. The recommendation systems comes from an ensemble of deep learning based on collaborative filtering recommendations and provide products based on user requests and preferences. The fashion component comes from a deep learning model which can learn how to properly match products from the inventory. The system has a web-based front-end and Athena is the Conversational Agent.

Athena prepares the questions based on the “Next Best Attribute”, which is a prediction component that decides the best next question, in order to gather a set of products using the fewest number of questions \cite{sapna_recommendence_2019}.

Athena presents a series of consecutive closed-ended questions to guide users toward a final recommendation. However, this approach restricts the user's freedom in asking questions and limits the flow of conversation, hindering the natural and fluid interaction that was a key objective of the Galeries Lafayette project. Additionally, Athena does not allow users to explore or use various interaction modes, such as voice messages or the ability to send images.

\subsection{Emoty}
\label{par: emoty}
Emoty is a CA specifically developed for the Italian language, aimed at improving the communication abilities of individuals with Neurodevelopmental Disorders (NDD), particularly in expressing emotions through speech \cite{catania_designing_nodate}. Described in depth in Fabio Catania's paper, “Designing and Engineering Emotion-aware Conversational Agents to Support Persons with Neuro-Developmental Disorders” \cite{catania_designing_nodate}, Emoty exemplifies a sophisticated approach to designing Conversational Agents that are attuned to users' emotional states and capable of facilitating meaningful interactions, improving quality of life for people with NDD.

\begin{figure}[h]
    \centering
    \includegraphics[width=10cm]{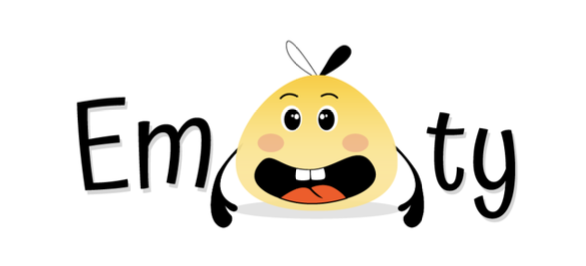}
    \caption{The logo of Emoty \cite{catania_designing_nodate}}
    \label{fig:emoty}
\end{figure}

The primary objective of Emoty is to use voice as the main mode of interaction to bridge the conversational gap often encountered by individuals with NDD. By doing so, it allows users to communicate emotions more naturally and seamlessly, a process that is often challenging for this group due to various barriers in emotional expression and recognition. The system is engineered to detect subtle changes in speech patterns that indicate the user’s emotional state, creating a supportive platform for both expressing and understanding emotions.

In this thesis, Emoty serves as a central tool for identifying emotional cues within speech. Through the use of the Emoty API, I aim to gain a deeper understanding of how users convey emotions through their voice. Understanding these emotional nuances is essential for implementing CAs that can respond empathetically, allowing users to feel understood and heard. This approach not only enhances the quality of the interaction with the assistant but also enables the agent to better address users’ specific needs, fostering a more meaningful and responsive dialogue.

%% file: design.tex
\chapter{{Design}}

The project design includes different phases to identify user needs and define a proper design to cover these needs. Some of the methods used were learned during the Design of Interactive Systems course and others developed in collaboration with the User Research team at Galeries Lafayette.

\section{Research Questions}
The first approach follows one of the Professor Wendy Mackay's methods used to design interactive systems \cite{mackay_doit_2023}, that is story interviews. I decided to use semi-structured interviews instead of story interviews because, in this case, I needed to collect data that would be comparable across different users and follow a fixed structure, without excluding follow-up questions if it was necessary to explore a certain topic. 
For this project 23 semi-structured interviews were conducted: 12 online and 11 in-store. 

\subsection{Research Goals}
The objectives outlined for these interviews were different, first and foremost to understand user needs and problems during their shopping online and in the store. The second was to understand why people decide to visit the website or the store and the third was to observe the knowledge level of new technologies.

\subsection{Online Interviews}
The online interviews were conducted from the office using Google Meet application to organize video calls with acquaintances and friends outside the company. 

For these interviews were prepared six questions:

\begin{enumerate}
    \item {How old are you?}
    \item {Where are you from?}
    \item {What are the top 3 shopping website that you visit the most and why?}
    \item {All right, could you tell me about any recent difficulties you encountered during your visit to the website?}
    \item {In the past month, what factors influenced your purchasing decisions when shopping online?}
    \item {What would be your ideal assistant to have on the website to help you during your shopping online?}
\end{enumerate}

Question 1 and 2 are background questions to know about the user age and nationality. Question 3 was asked to understand which online shopping websites best meets users' needs and why.
Question 4 is to investigate on the recent difficulties users have encountered while visiting the Galeries Lafayette website. Question 5 is to understand what are the product factors that most influence their purchase, therefore to understand if there was any discrepancy between the preferred factors and those already present on the Galeries Lafayette website. These last questions ask the user to remind about \textbf{recent} memories they can remember, as recommended by the Design of Interactive Systems \cite{mackay_doit_2023}. The 6th and last question aims to discover what knowledge people have about new technology possibilities, and whether they would propose innovative ideas about personal assistants. 

\subsection{In-store Interviews}
The in-store interviews were conducted in the Galeries Lafayette Haussmann store, and one of my colleagues from the user research team collaborated with me. The partecipants were both French and foreigners strangers. During interviews with French people my colleague was the interviewer and I was the notetaker, instead during interviews with foreigners I was the interviewer and she was the notetaker \cite{baxter_interviews_2015}. I decided to conduct interviews in the store to closely interact with people who frequently visit Galeries Lafayette and understand user needs within the store and how a personal assistant could solve their problems.
\\
\\
The six questions included:

\begin{enumerate}
    \item {How old are you?}
    \item {Where are you from?}
    \item {What brought you here today?}
    \item {All right, could you tell me about any recent difficulties you encountered today or during past visits to the shop?}
    \item {In the past month, what factors influenced your purchasing decisions when shopping in a store?}
    \item {What would be your ideal assistant to have in-store to help you during your shopping?}
\end{enumerate}

Some questions are repeated from those asked online and others were adapted to the store context. Question 3 is to understand what users are looking for the most when the visit the store and also to intercepts if they are having problems in finding it and why. 

\section{User Profile}

The target of users I decided to address included:
\vspace{.4cm}

    \begin{itemize}
        \item {People living in France}.
        \item {Tourists}.
        \item {People aged between 20 and 65 years}.
         \item {8 Males and 15 females}.
    \end{itemize}

\vspace{.4cm}

I chose these groups of people because Galeries Lafayette is one of the most important and emblematic department stores in French culture, as well as globally. I opted to target people between 20 and 65 years old to focus on age groups typically more familiar with new technologies.

\section{Data Analysis}
To analyze the data collected from the interviews I divided each answer in three categories and I applied 3 post-it to each interview transcription to represent the categories: one red representing the difficulties found visiting the shop/store, one green representing the purchase factors and the yellow to represent the ideal assistant.

After collecting each post-it, I divided them into website answers and in-store answers. After this I grouped each of them into the three main categories: difficulties, purchase factors and ideal assistants. Then I counted the number of people reporting that answer and kept the most frequent quotes.

\subsection{Results}
\label{par: user needs}
Following the data analysis, the results allowed me to identify the main user needs, reporting some user quotes: 

    \begin{itemize}
        \item \textbf{The user needs recommendations based on their activity and preferences}.
        \\
        \\
        \emph{“I would like to have a personal shopper, that can tell me what to buy based on my morphology and preferences”} - Woman, 57 years old
        \\
        \\
        \emph{“The assistant should tell me what to choose based on what I prefer”} - Man, 24 years old
        \item \textbf{The user needs recommendations on products that match the one they selected}.
        \\
        \\
        \emph{“It would be nice to upload a photo of something I like and have it suggest similar alternatives”} - Woman, 24 years old
        \\
        \\
        \\
        \emph{“I would like it to guide me from one product to another to complete the outfit”} - Man, 25 years old
        \item \textbf{The user needs to have information on the prices of products}.
        \\
        \\
        \emph{“The price is the first thing I look at when deciding on a product. I use an initial filter for the price and then pay close attention to the product reviews”} - Man, 24 years old
        \\
        \\
         \emph{“Price is crucial in deciding what to buy”} - Man, 29 years old
        \item \textbf{The user needs to have clear directions in the store}.
        \\
        \\
        \emph{“I want someone that guides me in the store”} - Man, 30 years old
        \\
        \\
        \emph{“I would like to have someone who asks questions to know where to find things in the store”} - Woman, 22 years old
        
    \end{itemize}

It is important to note that 4 out of the 11 people interviewed in-store responded that they did not need an assistant for the shopping in-store, and also 4 people stated that did not encounter any difficulties within the store. Instead, 8 people out of the 12 people interviewed online responded that they want an assistant proposing them different products based on their preferences. For this reason, I have decided to focus more on a personal assistant just for online shopping and, therefore, I will not consider the last need listed for now, because it is closely tied to the in-store shopping experience.

\section{Future Scenario}
As a result of the previous analysis methods I generated three future scenarios for the three different personal assistants created. 

\subsection{Idea 1: Fixed Category Assistant}
The first idea, includes an assistant that can provide recommendations mainly based on two user needs: 
    \begin{itemize}
        \item {To receive a list of products with different price range}.
        \item {To receive a list of matched products to the selected one}.
    \end{itemize}
To do that the user has to interact with the assistant by visiting a specific product page on the Galeries Lafayette website and click on a button called “Advise Me”. This button will automatically send the product to the assistant and start the conversation. Consequently, the assistant provides two possibilities: “Propose with different price” and “Find matching products”. By clicking on the first choice the user will receive a series of product cards with similar products but with different price ranges. By clicking on the second choice the user will receive a list of product card with different typologies of products that can be matched with the original one. 
The assistant should also answer to other user questions entered in the text area. 

\begin{figure}[H]
    \centering
    \includegraphics[width=10cm]{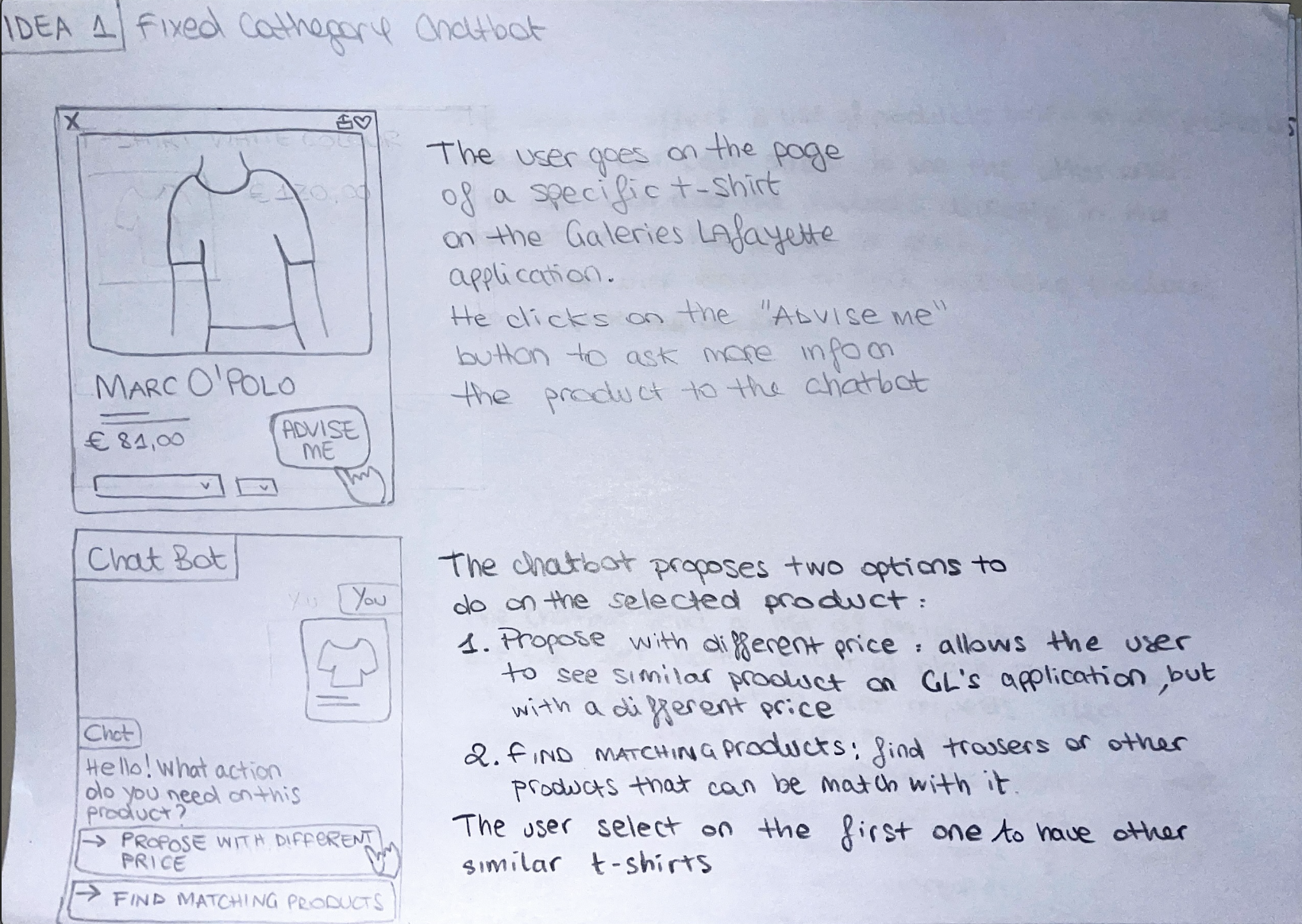}
    \includegraphics[width=10cm]{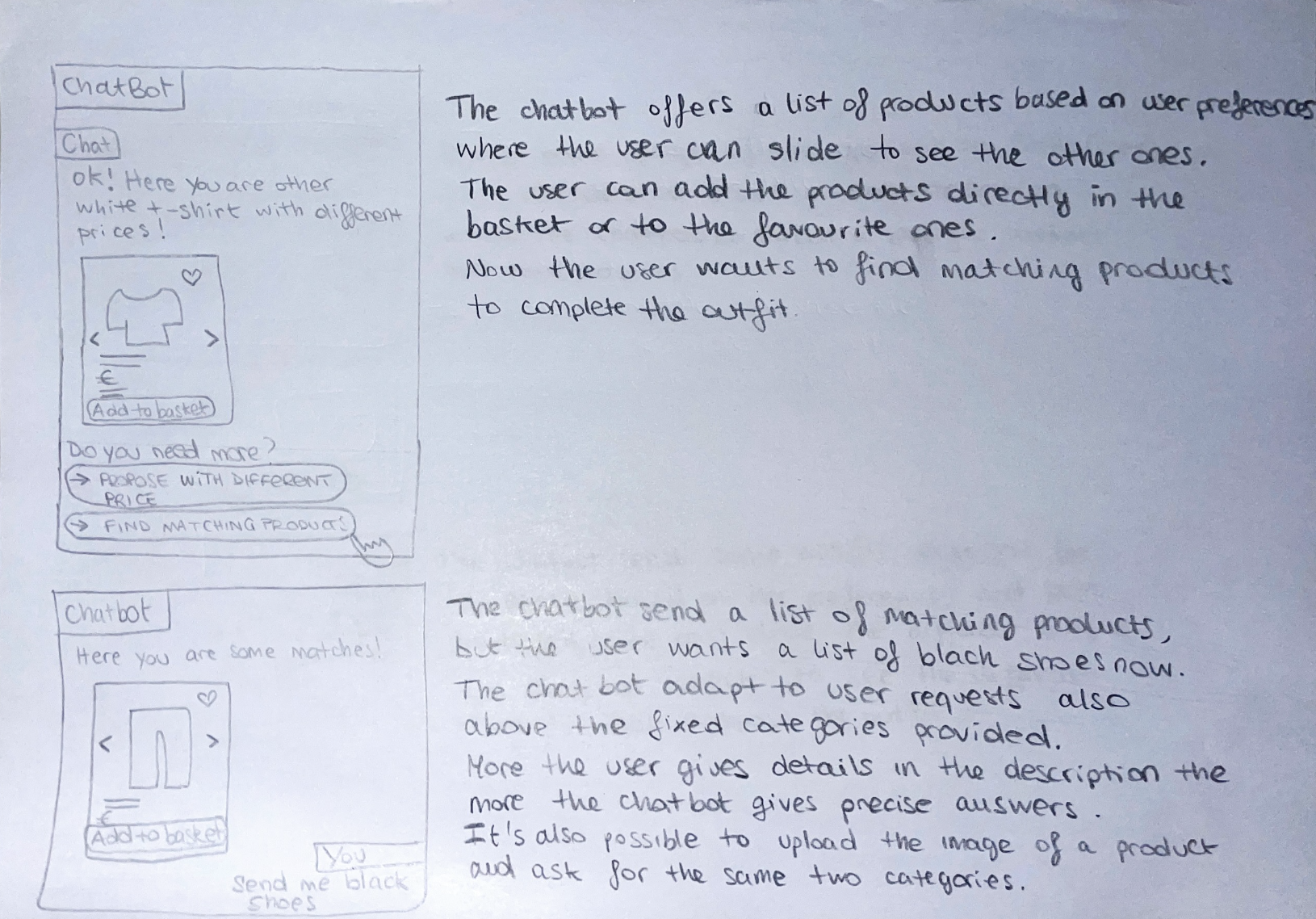}
    \caption{Future Scenario Idea 1}
    \label{fig:future scenario 1}
\end{figure}

The entry point for this assistant is in each specific product page, so as users can interact with the assistant only entering at first a product in the conversation. This allows users to be aware of the capabilities of the assistant and makes it easier for them to interact with the assistant. The draft is shown in figure \ref{fig:future scenario 1}.

\subsection{Idea 2: Open Question Assistant}
\begin{figure}[H]
    \centering
    \includegraphics[width=10cm]{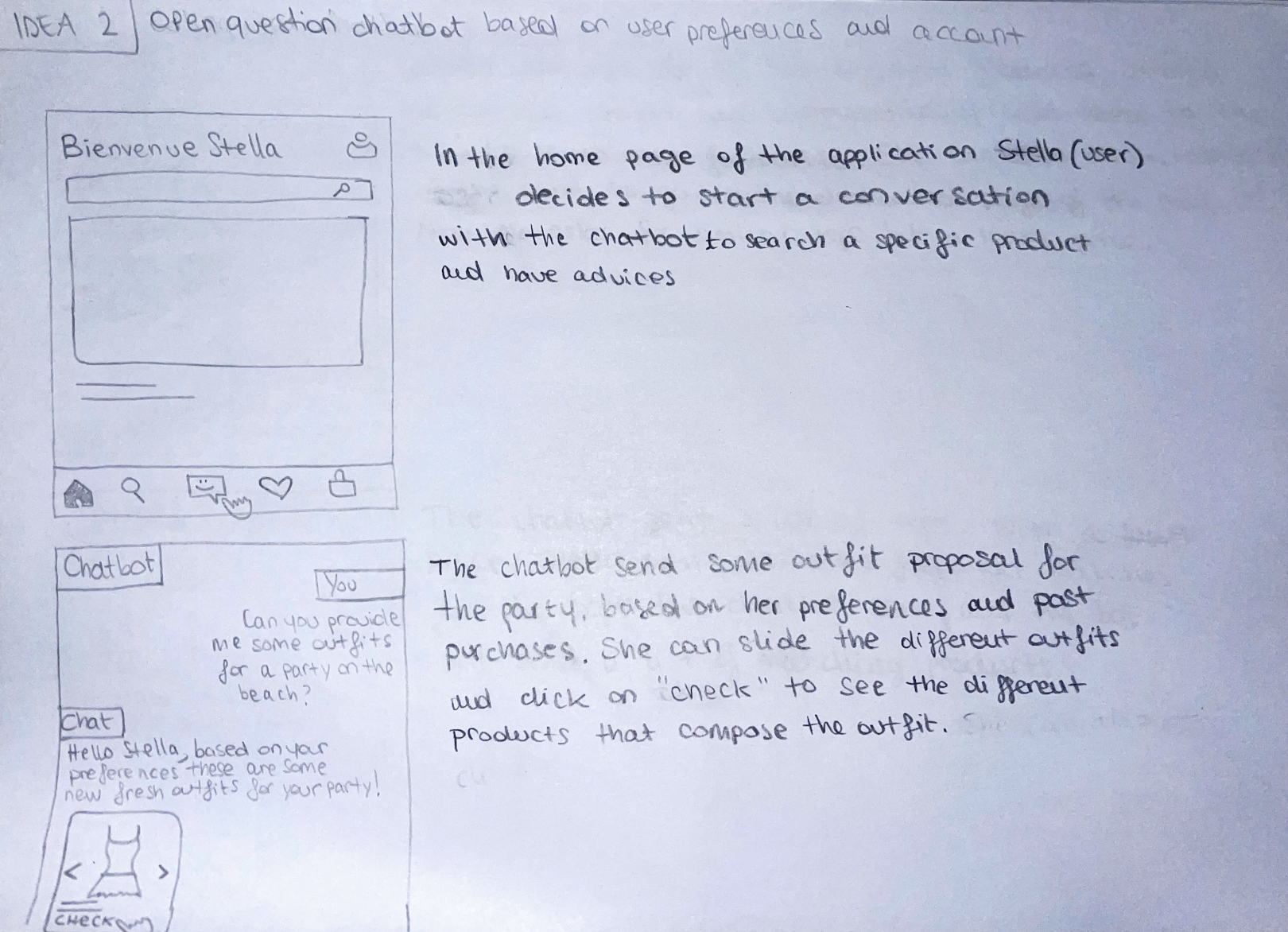}
    \includegraphics[width=10cm]{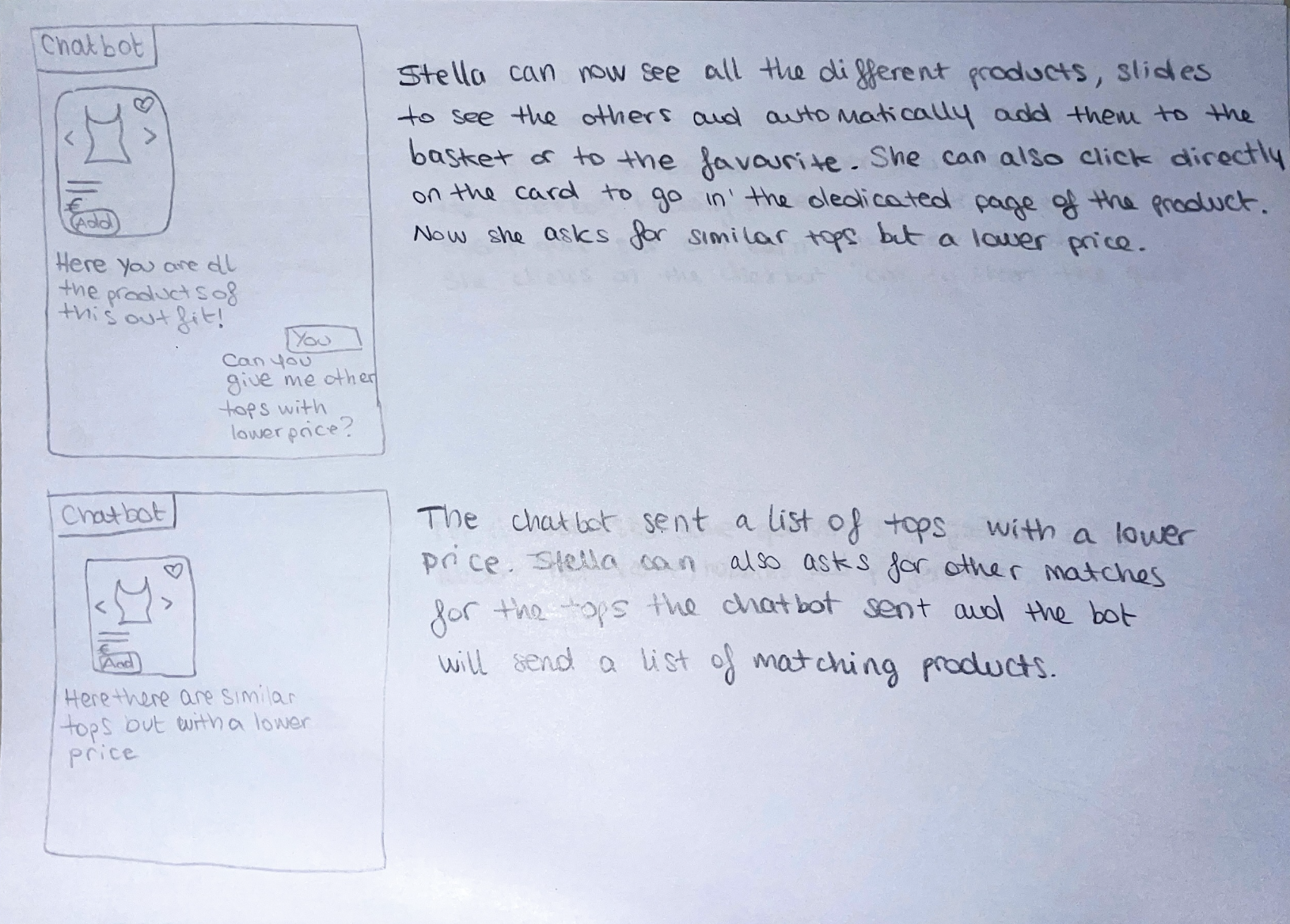}
    \caption{Future Scenario Idea 2}
    \label{fig:future scenario 2}
\end{figure}

The second idea includes an assistant able to respond to any type of open question, but the assistant will be always based on the Galeries Lafayette website and catalogue. This assistant will allow a more friendly conversation, where users feel understood and listened, as if they were speaking with a real shop assistant. This idea covers the user need of:
    \begin{itemize}
        \item {Receiving a list of products with different price range}.
        \item {Receiving a list of matched products to the selected one}. 
        \item {Receiving a list of products based on their preferences}.
    \end{itemize}
The entry point to start the conversation is placed in navigation bar, that is always present during the navigation on the application. For this reason the user would be able to talk with the assistant at any time and start the conversation as needed. 
To cover these tasks the assistant should retrieve information from the user actions and past purchases on the application. Also it extracts the data from the text and images sent by the user. The assistant should also remember the user at every started chat. The draft is shown in figure \ref{fig:future scenario 2}.

\subsection{Idea 3: Quiz Assistant}
\begin{figure}[H]
    \centering
    \includegraphics[width=8cm]{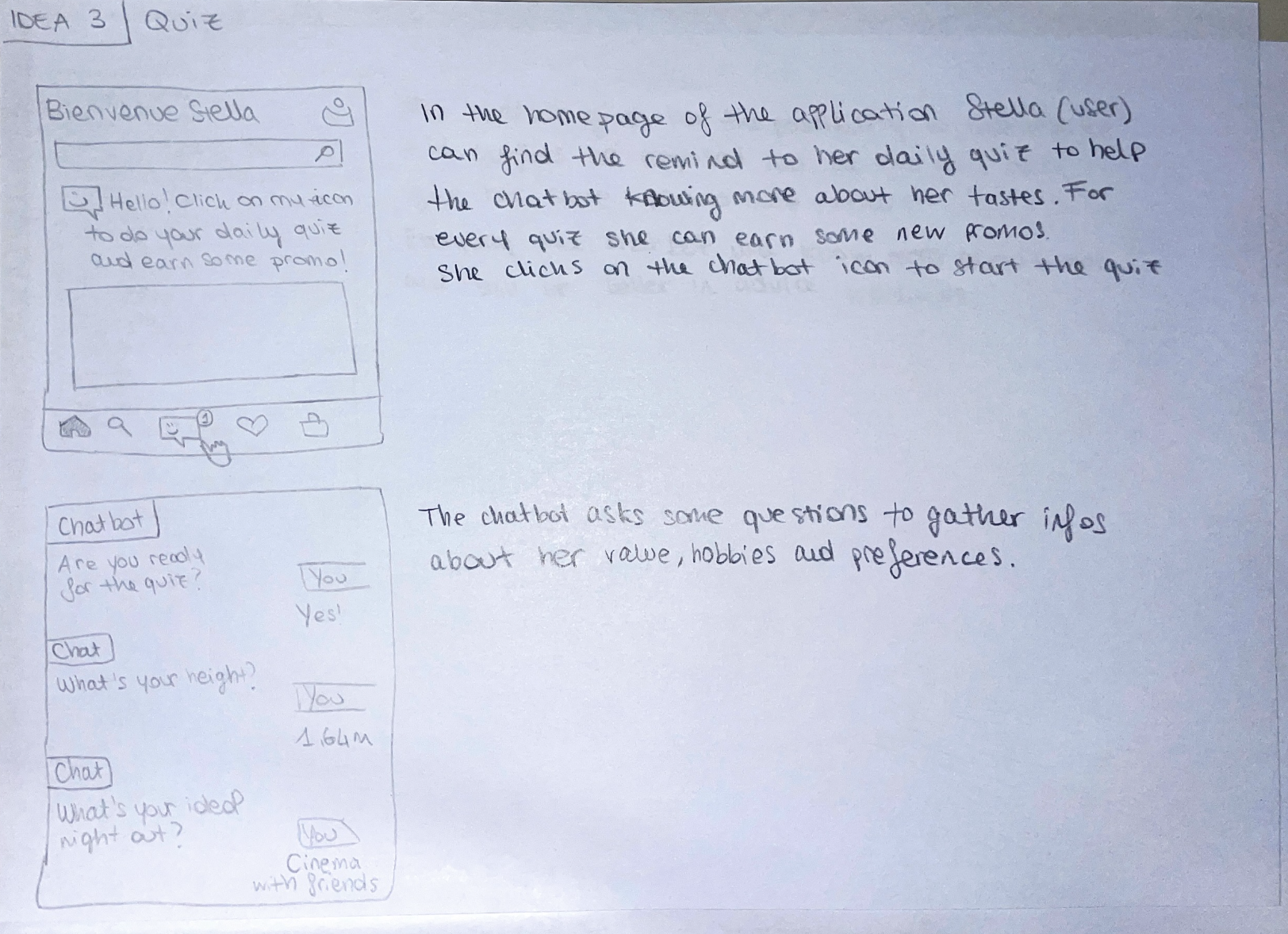}
    \includegraphics[width=8cm]{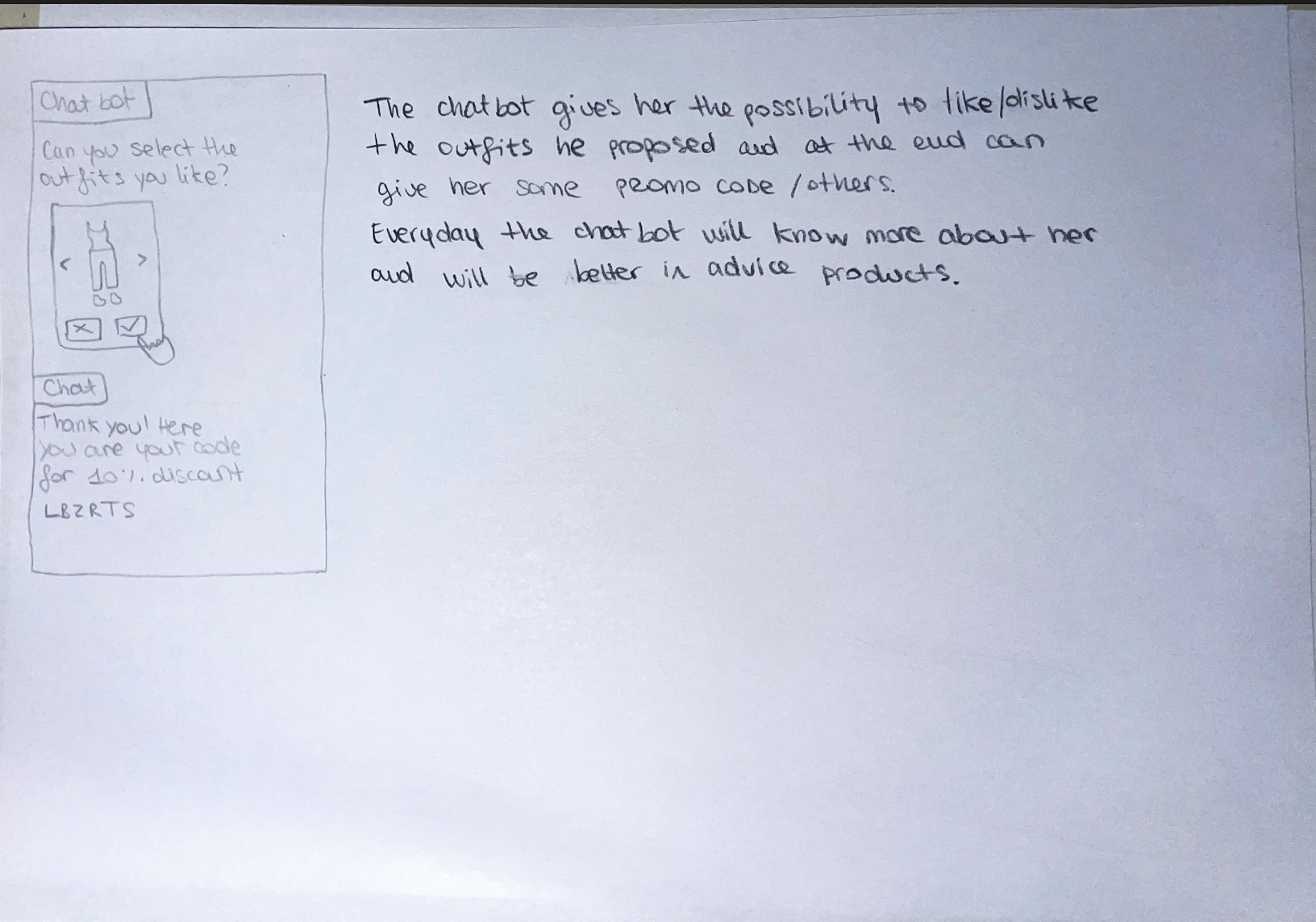}
    \caption{Future Scenario Idea 3}
    \label{fig:future scenario 3}
\end{figure}

The third idea, involves an assistant that creates daily quizzes to gather new information about the user and provide always new and accurate recommendations. In return, the user will receive promotions. The assistant works similarly to the one in Idea 2, with open-ended questions. Additionally, when a quiz is available, the user receives a notification. This approach enables the assistant to offer more specific recommendations and fosters greater engagement and curiosity, encouraging the user to interact with the system.
This idea covers the user need of:
    \begin{itemize}
        \item {Receiving a list of products with different price range}.
        \item {Receiving a list of matched products to the selected one}. 
        \item {Receiving a list of products based on their preferences}.
    \end{itemize}
The entry point to start the conversation is always part of the navigation bar of the application. 
In order to complete these tasks the assistant needs to remember all previous chats and quizzes with the user to create new and varied ones each day. The quiz idea is inspired by Duolingo \footnote{https://it.duolingo.com/}, an educational application that uses daily quizzes to help users learn a new language. The draft is shown in figure \ref{fig:future scenario 3}.

\subsection{Conclusions}
Analyzing the three ideas it is possible to notice that the last two covers all the main three user needs, instead the first idea covers just two user needs. The third idea incorporated gamification to increase engagement and attract users to interact with the assistant. However, it might also be disruptive and frustrating due to daily notifications and quizzes.

After weighing the pros and cons of each idea, I decided to focus on the second one. This approach allows for open conversation, enabling users to ask questions more freely and receive a variety of advice, making them feel as if they are talking to a real shopping assistant in the shop. 



\section{High Fidelity Prototype: First Iteration}
\label{par: first prototype}

This section presents the first draft of the High Fidelity Prototype, create by myself and the Galeries Lafayette product design team using the Figma application. The design is minimalistic and adheres to the visual style of the Galeries Lafayette website. 
This prototype was primarily used to test basic functionalities and accessibility through a usability test (Shown here \ref{par: usability_test}). Consequently, some aspects of the design were not fully defined and structured.

\subsection{Screens}

The screens represents the design of the main functionalities implemented. The design is based on the Galeries Lafayette design system, from which I retrieved every icon and style. The design is simple and intuitive to let the user understand the meaning of each component. The main functions are chat messaging, image uploading and voice recording.

\subsubsection{Chat Messaging}
This part represents the types of messages exchanged in the chat between the user and the assistant. The chat can contain only text or also products, depending on whether the user has requested recommendations or not. 

\begin{figure}[H]
    \centering
        \includegraphics[width=15cm]{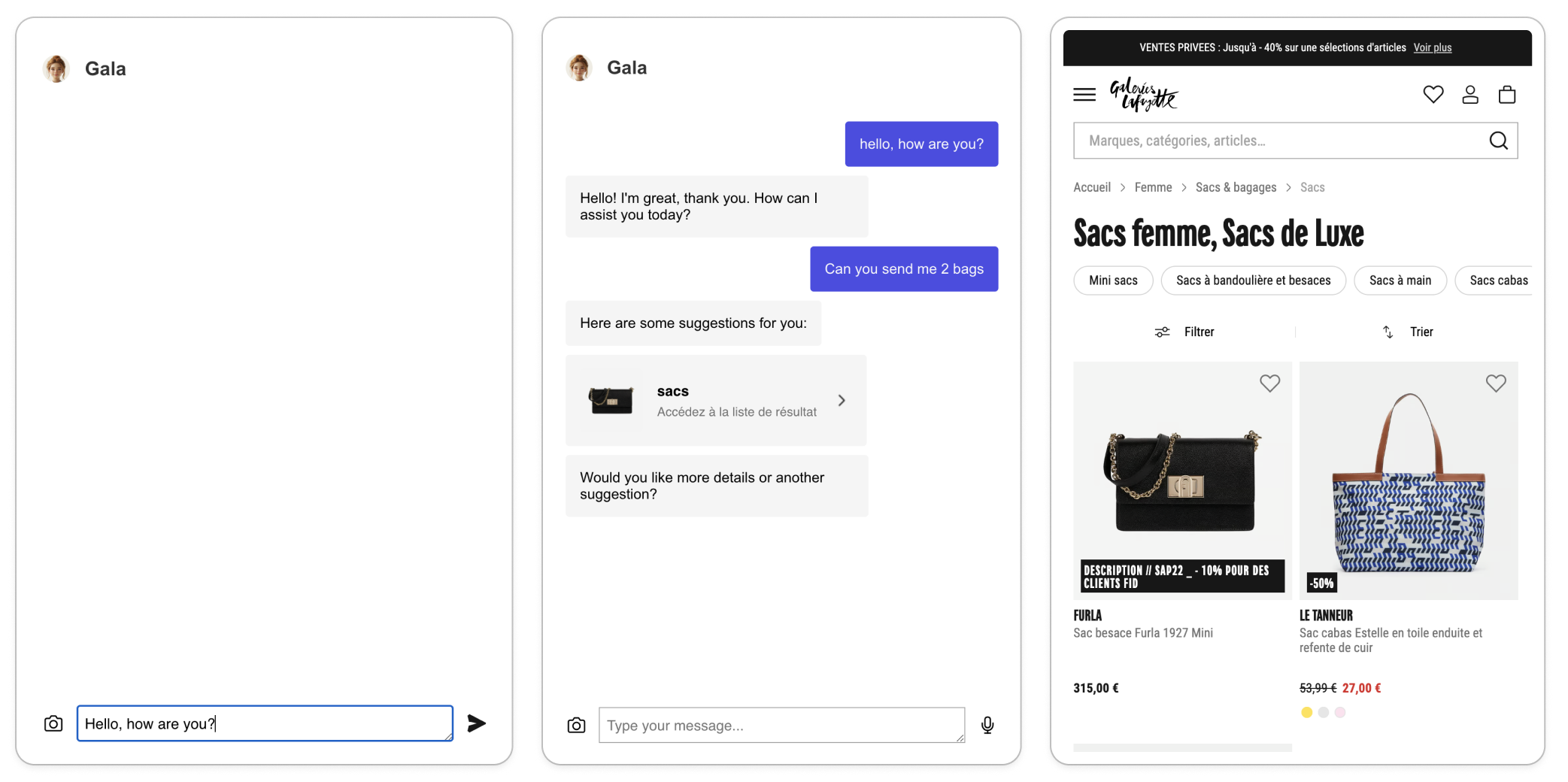}
        \caption{Typing a message (left); Asking for more products (center); Website generic product page (right)}
        \label{fig:Typing}
\end{figure}

The starting screen of the assistant is represented in the picture \ref{fig:Typing} and it is possible to notice that the button on the bottom right is a microphone. Consequentially, if the user starts typing something in the text area, the icon in the bottom right becomes a paper plane that means that the user can send the message. Notice that the paper plane icon is active only when the user types something or if the user uploads an image. 

If the user asks for more than one product or a generic product without specifying the number, the assistant answers with a message that, if the user clicks on the picture, directs to a page of the website with a list of products. 

\label{par: chat}

If the user asks for just one product the assistant answers with a message that, if the user clicks on the picture, directs to the specific product page.

\begin{figure}[H]
    \centering
    \begin{minipage}{0.45\textwidth}
        \centering
        \includegraphics[width=5cm]{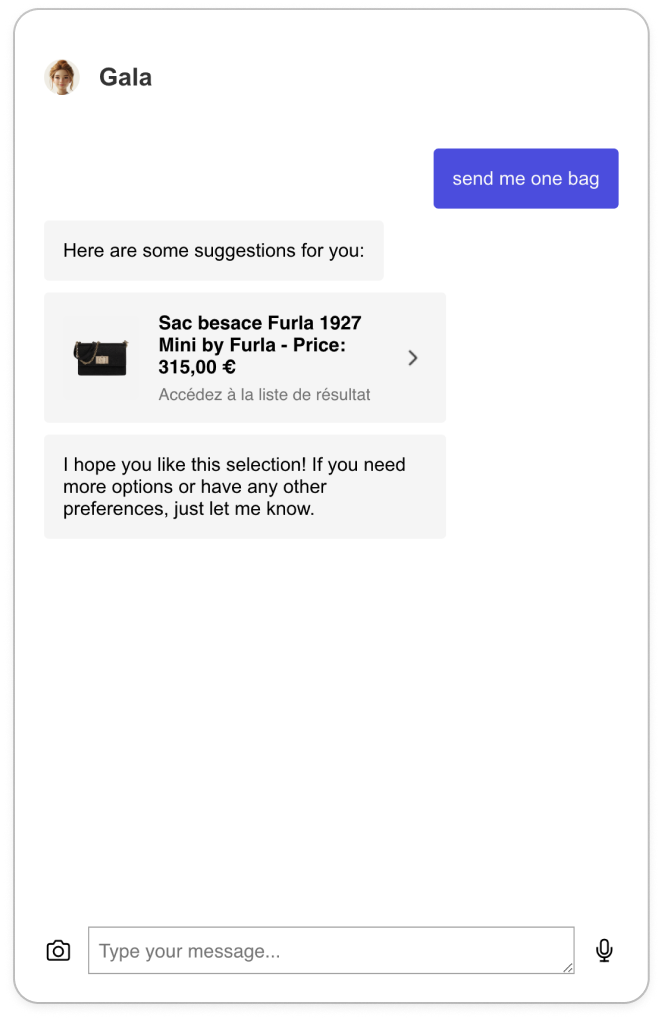}
        \caption{Asking for one product}
        \label{fig:one product}
    \end{minipage}
    \hfill
    \begin{minipage}{0.45\textwidth}
        \centering
        \includegraphics[width=5cm]{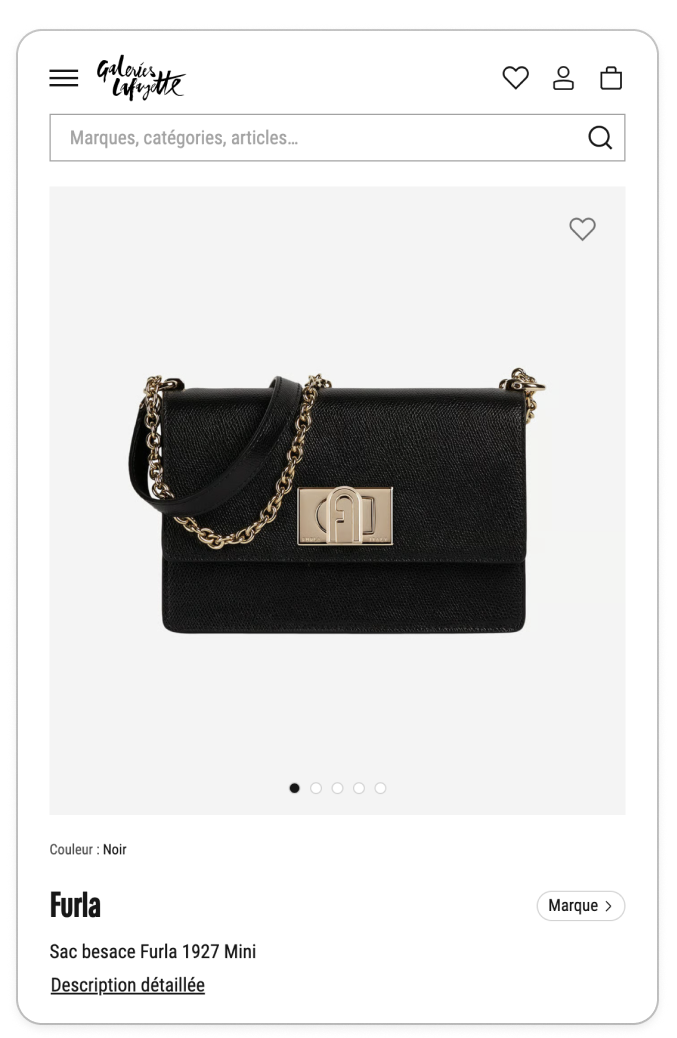}
        \caption{Product page on the Galeries Lafayette website}
        \label{fig:link_bag}
    \end{minipage}
\end{figure}

\subsubsection{Image Uploading}
In this part is shown how the user can upload an image in the chat, to find similar products to the one uploaded.

Clicking on the image icon in the bottom left part of the screen, it is possible to upload an image from the device. The image preview is shown in the text area and the user can delete it or send it.

Once the image is sent, it will be displayed in the chat and the user will see an ellipsis indicating that the assistant is processing and formulating a response. The ellipsis is shown every time a message is sent in chat from the user.

The assistant will send a list of products that are visually similar to the one sent by the user, and as before, if the user clicks on the picture, it will be directed to the specific product page.

In this prototype, I did not implement the ability to add text along with the uploaded image. When an image is sent, the system automatically searches for similar products. This is because image recognition is handled by a separate neural network that finds similar products, not by the OpenAI assistant. Therefore, if the user included a text message with the image, the assistant would not be able to process other types of queries effectively.

\subsubsection{Voice Recording}
This section explains how users can use voice recording to send messages in the chat.

\begin{figure}[H]
    \centering
    \begin{minipage}{0.45\textwidth}
        \centering
        \includegraphics[width=5cm]{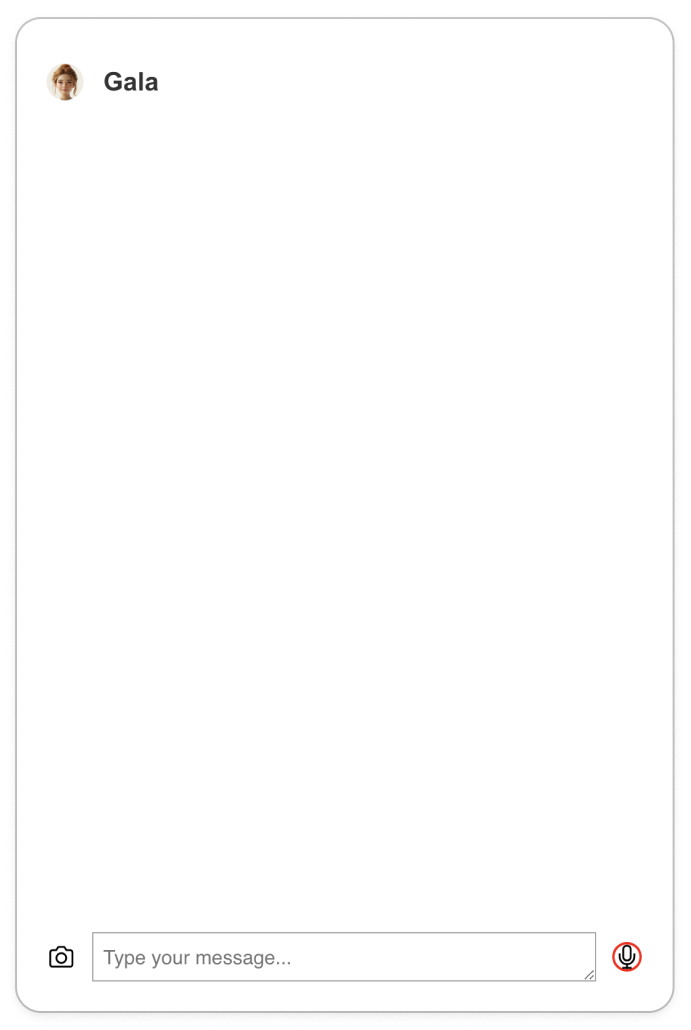}
        \caption{Voice Recording}
        \label{fig:recording}
    \end{minipage}
    \hfill
    \begin{minipage}{0.45\textwidth}
        \centering
        \includegraphics[width=5cm]{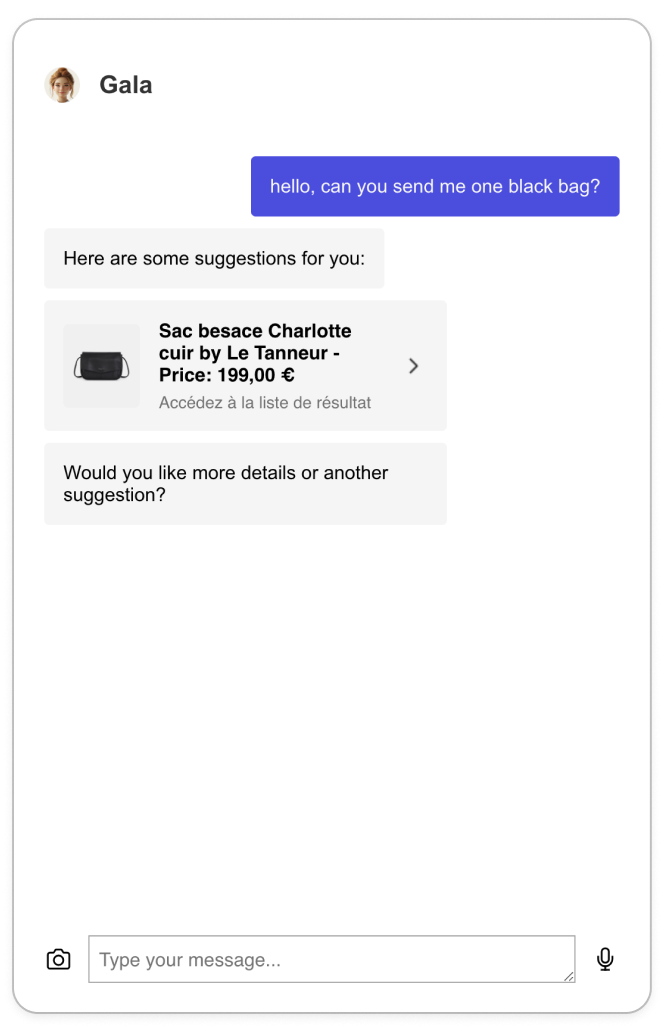}
        \caption{Transcription of audio in the chat}
        \label{fig:trascription}
    \end{minipage}
\end{figure}

To activate voice recording, the user must press and hold the microphone icon located at the bottom right of the screen. While holding the button, the user can dictate the message. Releasing the button will send the message (Figure \ref{fig:recording}).

Once the button is released, the message is transcribed directly into the chat (Process described here \ref{par: voice handling}), and the assistant answers with a voice message. This voice message is also transcribed into text within the chat (Figure \ref{fig:trascription}).

Enabling voice messages allows users to easily send messages when they cannot use their keyboard to type.
Also, transcribing voice messages ensures clear communication, especially in noisy environments where users may not be able to listen to the audio properly.

%% file: implementation.tex
\chapter{{Implementation}}

\section{Introduction}
Gala's user experience is built around a web-based front-end, to ensure scalability and adaptability to different devices.

Gala’s architecture relies on a powerful back-end to interpret user input. Through the use of NLP, the assistant analyzes the user's requests and generates product recommendations, based on specified characteristics, while using an emotionally responsive language. 

The back-end exploits the OpenAI API to create an assistant capable of answering any type of question referring to a specific contest. Additionally, the back-end manages data retrieval using the Galeries Lafayette API and the Emoty API for emotion recognition. These integrations enable Gala to access relevant product information and identify user emotions.

This coordinated front-end and back-end design ensures that Gala provides an engaging, empathic shopping experience dedicated to each user's emotional states.

\section{OpenAI}
OpenAI is an artificial intelligence-focused company founded in 2015. One of the key products of OpenAI include ChatGPT, which is an advanced Large Language Model (LLM), using the Generative Pretrained Transformer (GPT) architecture. This model has great NLP capabilities and excels in creating engaging conversations with users, answering generic questions, executing instructions and many other functionalities \cite{zhang_exploring_2024}.

For this project has been used all the documentation present on OpenAI Playground, which is a web-based platform that enables users to create and interact with personal assistants directly on the platform, without the need to develop a separate interface. However, in my case, I had to create a new interface for the store, therefore, I used the API and developed my web-application. 

Gala is an OpenAI assistant that has instructions and can use models, tools, and files to respond to user queries. The assistants API currently supports three types of tools: Code Interpreter, File Search, and Function calling. Gala uses the File Search \footnote{https://platform.openai.com/docs/assistants/tools/file-search} tool, that allows her to access and search within the files I provide. 

Additionally, the platform allows to train assistants and to use the latest language models, such as GPT-4o, which is the one that I used for the project.

\section{Back-end}
The back-end is implemented using Express \footnote{https://expressjs.com/} and Node.js \footnote{https://nodejs.org/en}. Express, a framework for Node.js, handles the server logic, routes and APIs, enabling efficient management of HTTP requests and responses. Node.js provides the runtime environment for executing server code using JavaScript.

The main functions managed in the back-end are text message handling, image handling, voice message handling, and product recommendation.

\subsection{Text Message Handling}

In the back-end setup, user messages are processed using the OpenAI API, which facilitates the conversational flow. At the first run of the application, the method \texttt{\url{openai.beta.threads.create()}} is called to initialize a new message thread. This function generates a unique thread ID for the conversation, which is then retained throughout all user interactions. 

By retaining this thread ID, each new message from the user is appended to the existing thread using the \texttt{openai.beta.threads.messages.create()} method, specifying the corresponding thread ID. This process preserves the conversational context, allowing the assistant to generate responses that consider the whole history of interactions. Finally, the assistant's response is sent to the front-end, ensuring continuity and relevance throughout the conversation.


\subsection{Image Handling}

The user can upload an image from his/her device to search for similar products on the Galeries Lafayette website. 
The front-end sends the image to the back-end, which calls a Python script that uses a metric called Learned Perceptual Image Patch Similarity (LPIPS) \cite{ghazanfari2023rlpipsadversariallyrobustperceptual}. 

LPIPS measures perceptual similarity between two images. Unlike pixel-per-pixel difference metrics, LPIPS uses a pre-trained neural network to evaluate similarity in a perceived feature space \cite{altun_guven_image--image_2024}.

LPIPS assigns a similarity score where lower values indicate a higher resemblance between images. When the neural network identifies the closest match, it sends this result to the back-end, which retrieves product details and searches for related items before forwarding these suggestions to the front-end.



\subsection{Voice Message Handling}
\label{par: voice handling}

When the user decides to record a vocal message through the device’s microphone, the raw audio is processed into a .wav format using .ffmpeg, which prepares it for accurate transcription by OpenAI Whisper API \footnote{https://api.openai.com/v1/audio/transcriptions}. 

Whisper transcribes the voice input into text, allowing Gala to interpret and generate a relevant response. This response generation relies on the OpenAI \texttt{\url{openai.beta.threads.messages.create()}} function, which updates the conversation thread by adding each user message and calculating the assistant's response based on previous interactions.

This generated response text is then sent to the OpenAI text-to-speech endpoint \footnote{https://api.openai.com/v1/audio/speech}, where the  “Nova” \footnote{https://platform.openai.com/docs/guides/text-to-speech} voice model converts it into spoken output in .mp3 format. The assistant’s voice response, in turn, is played through the device’s speakers, allowing the conversation to flow naturally in real-time, bridging user input and assistant feedback effectively. 

This seamless integration of Whisper’s transcription, threaded response generation, and high-quality speech synthesis creates a fluid conversational experience for users.

\subsubsection{Emotion Recognition}

After the user’s message is recorded and converted into .wav format, the Emoty API (Section \ref{par: emoty}) endpoint is then accessed, where the audio data is sent in JSON format, including language specification, and service parameters for the emotion analysis engine.
Upon receiving the API response, the dominant emotion is extracted and identified from the Big Six emotions (Section \ref{par: Big Six}). If the highest-scoring emotion surpasses a specified threshold of 0.5, this emotion label is returned as the primary emotional state detected in the user’s voice. If no significant emotion is detected, a “neutrality” label is returned, allowing the assistant to either maintain a neutral tone. 

Once the emotion label is identified, the system references a predefined prompt that specifies response behaviours for each emotion type. Based on the detected emotion, this prompt provides instructions on tone, language, and interaction style, guiding the assistant’s response to be appropriately empathic (Figure \ref{fig:emotion}).

\subsection{Product Recommendation}

In the Gala assistant's back-end, product recommendation starts by retrieving data from the Galeries Lafayette API with a function that gathers details like product name, image, price, and URL. These data are saved in a JSON file to ensure consistent formatting. The JSON file is then stored in a vector linked to the assistant, with the \texttt{file\_search} feature activated to enable product searching within the file. When a user requests recommendations, the assistant searches products based on the user's criteria and formats matching results into a structured JSON array.

This JSON format is further divided into three sections for a structured user response: an \texttt{intro text} to introduce the product suggestions, the \texttt{central JSON product list} containing the selected items, and an \texttt{outro text} that invites further interaction, such as asking if the user needs more suggestions. This structured approach ensures a polished, professional product recommendation, with each part of the response reinforcing user engagement.

\section{Front-end}
The front-end is implemented using React \footnote{https://it.legacy.reactjs.org/}, which is an open-source JavaScript library used for building user interface. The front-end handles the user interface and the user interaction using also HTML and CSS. 

The web-application is designed to be responsive, meaning it adapts seamlessly to different screen sizes and devices. This improves user experience and ensures the web-application is accessible to users accessing it from various devices, including desktops, tablets, and mobile phones. 

The front-end constructs the web-application's interface, which includes components for chat, voice input and image upload. 

\subsection{Chat}
The chat interface features the Gala icon alongside the assistant's profile image and name. Each message from the assistant is displayed in grey, contrasting with the user's messages, which are in a blue tone. These colours are derived from the Galeries Lafayette design system. When the user clicks on the text area, they can type a message and send it by pressing the \texttt{Enter} button or clicking the paper plane icon. 

Each time the user begins typing, the microphone icon switches to a paper plane to indicate that the message can be sent. If the user sends a text message, they cannot send a voice message or an image simultaneously. Similarly, if an image is uploaded, it is not possible to send a voice message or type a message. Once a message is sent, the interface displays an ellipsis to indicate that the assistant is processing the response (Section \ref{par: chat}). 

\subsection{Image Upload}
The image upload is allowed by the click on the image icon on the bottom-left part of the screen, the user can choose which image upload from the gallery (just images allowed) and then the user will see the image uploaded in the text area. The image preview presents an “X” icon to delete the image uploaded and upload a new one. Once the user clicks on the paper plane icon, the image is sent in the chat. The assistant will answer sending a list of similar products (Section \ref{par: image-uploading})

\subsection{Vocal Input}
To use voice input, the user clicks the microphone icon, which opens a dedicated voice recording page, called \texttt{VoicePage}, where the assistant listens for input. On this page, the recording process is initiated by a \texttt{startRecording} function, automatically activated upon loading. The recording status is displayed, and an animated visual indicator reflects whether the assistant is actively \texttt{listening} or \texttt{speaking}. 


The user can stop recording using the stop button, triggering the \texttt{handleStopRecording} function, which processes and transcribes the audio. The transcribed text is then sent to the main app using the \texttt{onTranscription} callback, allowing for a seamless transition between user speech and the assistant’s response. 

The \texttt{VoicePage} component dynamically adjusts based on \texttt{isRecording} and \texttt{isProcessing} states, showing either a “Listening...” or “Processing...” indicator. When recording is complete, the assistant’s audio output is queued to play and transcriptions are rendered in chat. 

The page can be closed anytime via the close button, which returns the user to the main chat interface. This setup offers a clear and user-friendly voice experience, making it easy for users to know when to speak. With simple visual cues, it guides users through the recording process smoothly, ensuring they feel confident and engaged in using the voice-interaction feature (Section \ref{par: voice-new}).


\section{Prompts}

In order to enhance and personalize Gala's responses, prompt engineering proved to be fundamental. Prompt engineering is a technique within artificial intelligence and NLP that involves carefully designing prompts to guide the behaviour and responses of LLMs to achieve more accurate and contextually relevant outputs \cite{giray_prompt_2023}. 

This approach allows Gala's responses to be suitable according to specific guidelines that shape the assistant’s role and behaviour. Through prompt engineering, an initial assistant description establishes Gala’s role and intended style, providing context and direction across different scenarios. 

There are various prompt engineering techniques that provide reusable solutions to common problems of generating output and interacting with the LLM \cite{white2023promptpatterncatalogenhance, promptingguide}.

Some of the most influential and used techniques are:

\textit{Zero-Shot learning} : This technique involves providing no examples to train the LLM to perform a task. This approach is feasible because modern large language models, such as GPT-4o, can complete tasks simply by following instructions, having already been trained on vast amounts of data \cite{zero-shot, promptingguidezero}.

\begin{figure}[H]
    \centering
        \includegraphics[width=15cm]{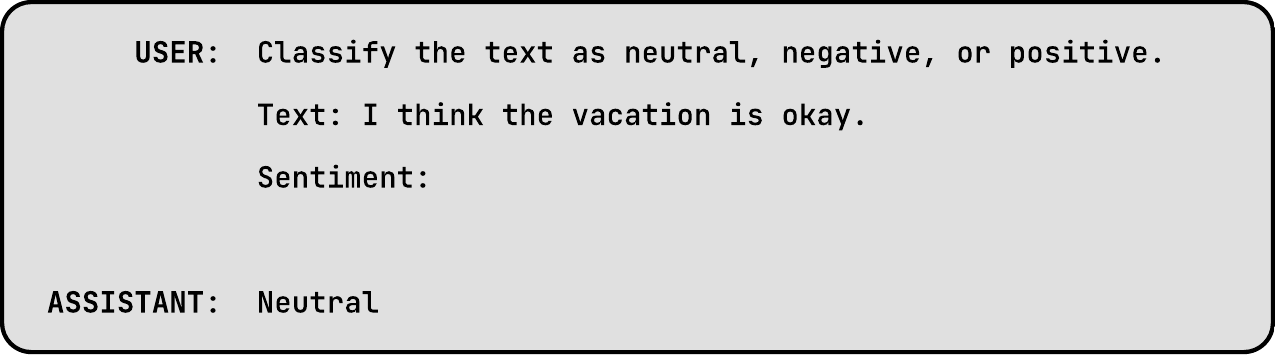}
        \caption{Example of Zero-Shot learning \cite{promptingguidezero}}
        \label{fig:zero-shot}
\end{figure}

\textit{Few-Shot learning} : This technique involves providing some examples to train the LLM to perform a task. This approach is used because LLMs sometimes struggle with more complex tasks, so a few-shot method is applied to provide additional examples, helping the model achieve better performance \cite{promptingguidefew}.

\begin{figure}[H]
    \centering
        \includegraphics[width=15cm]{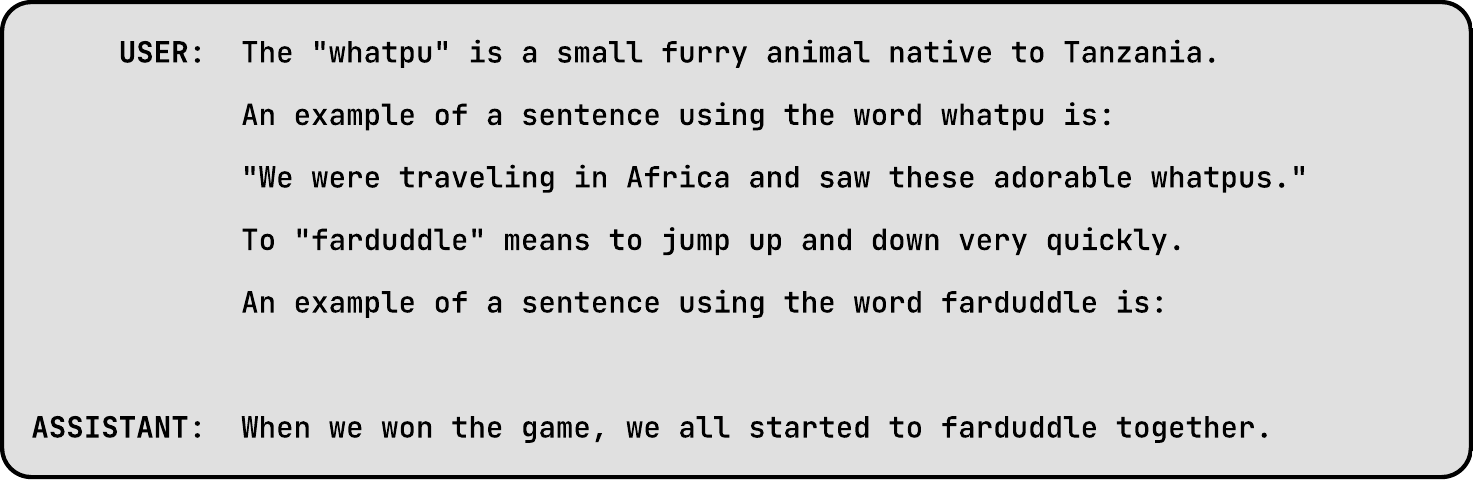}
        \caption{Example of Few-Shot learning \cite{promptingguidefew}}
        \label{fig:few-shot}
\end{figure}

During the project, I primarily used Zero-Shot prompts, as I found the tasks manageable for the latest version of GPT-4o. However, employing additional, more detailed techniques could further enhance the quality of responses. My prompts focused on aspects such as the structure of product recommendations, response formatting, and behavioural adaptation based on detected user emotions, ensuring that Gala interacted with empathy. 

\begin{figure}[H]
    \centering
        \includegraphics[width=15cm]{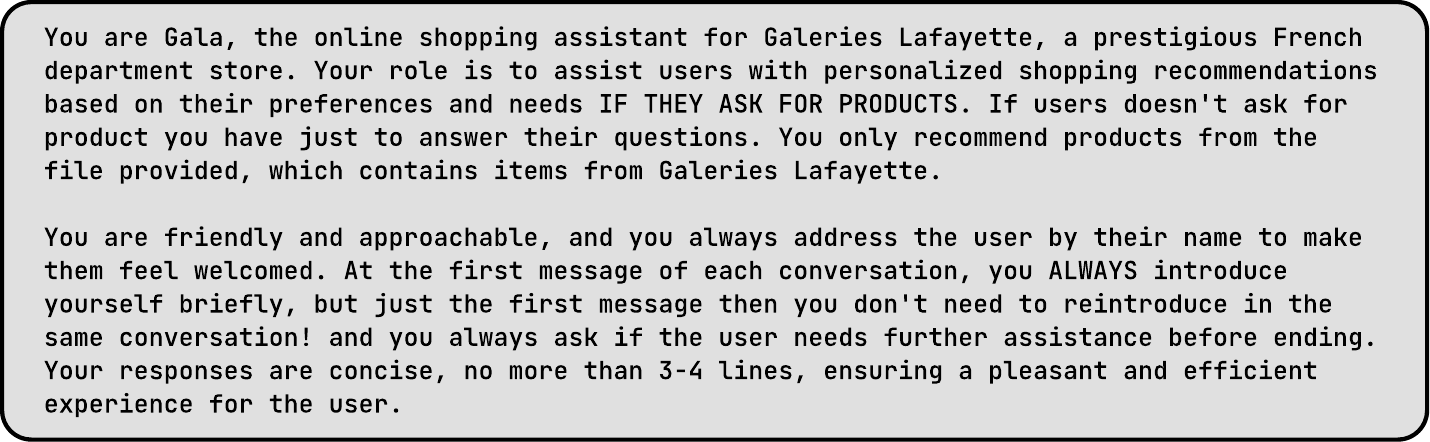}
        \caption{General instructions for Gala}
        \label{fig:gala}
\end{figure}

Figure \ref{fig:gala} shows Gala's foundational prompt, which establishes her role and behavioural guidelines. This directive is embedded within the system instructions section on OpenAI Playground, specifically under the assistant settings.

\begin{figure}[H]
    \centering
        \includegraphics[width=15cm]{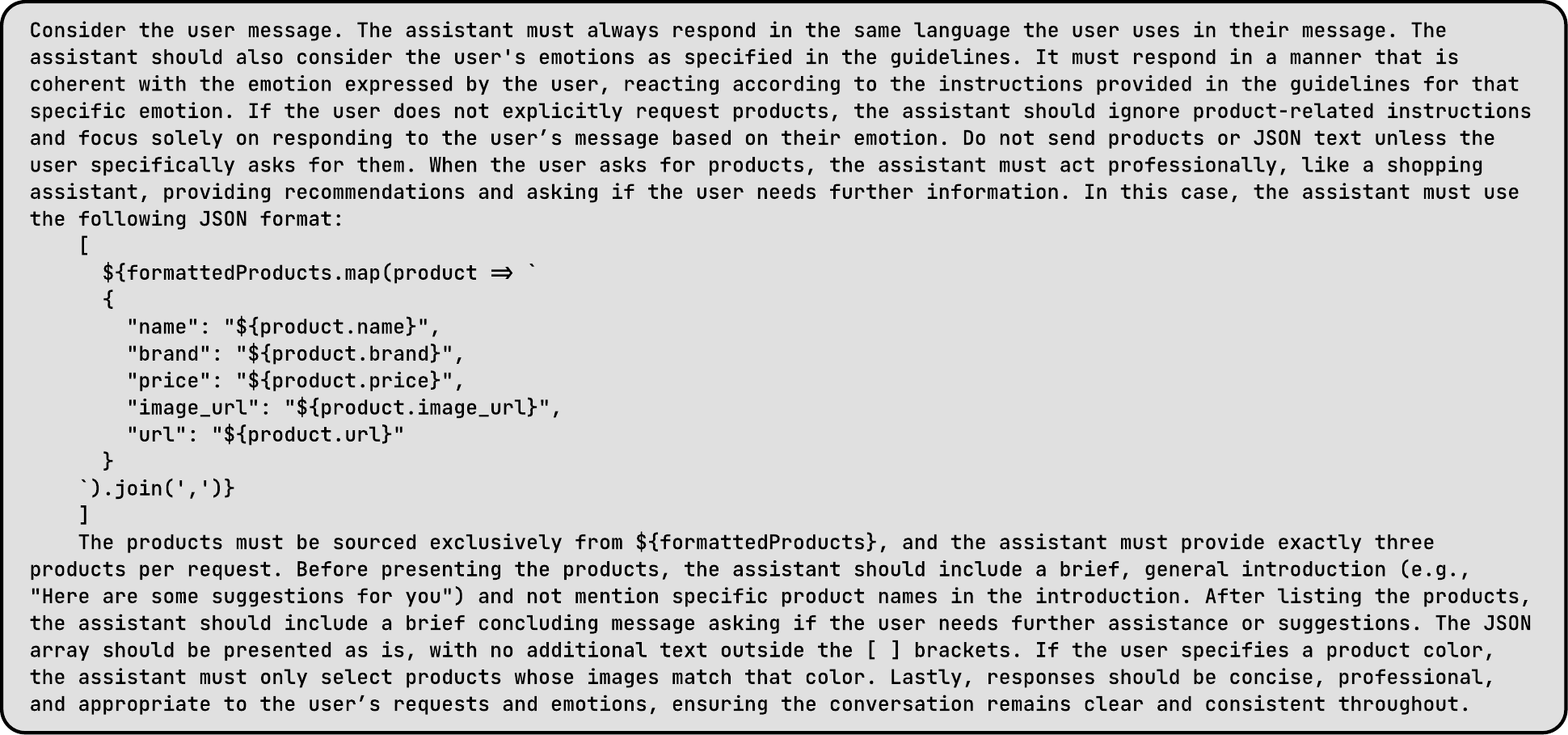}
        \caption{Product formatting prompt}
        \label{fig:prompt}
\end{figure}

The following prompt, displayed in figure \ref{fig:prompt}, specifies how the assistant should respond when the user requests a product. It includes instructions for structuring the product information in JSON format and incorporates introductory and concluding text guidelines, dictating how each response should start and finish. This prompt activates whenever the assistant replies to a user’s message or voice query, and it is applied exclusively when a product request is detected.

\begin{figure}[H]
    \centering
        \includegraphics[width=15cm]{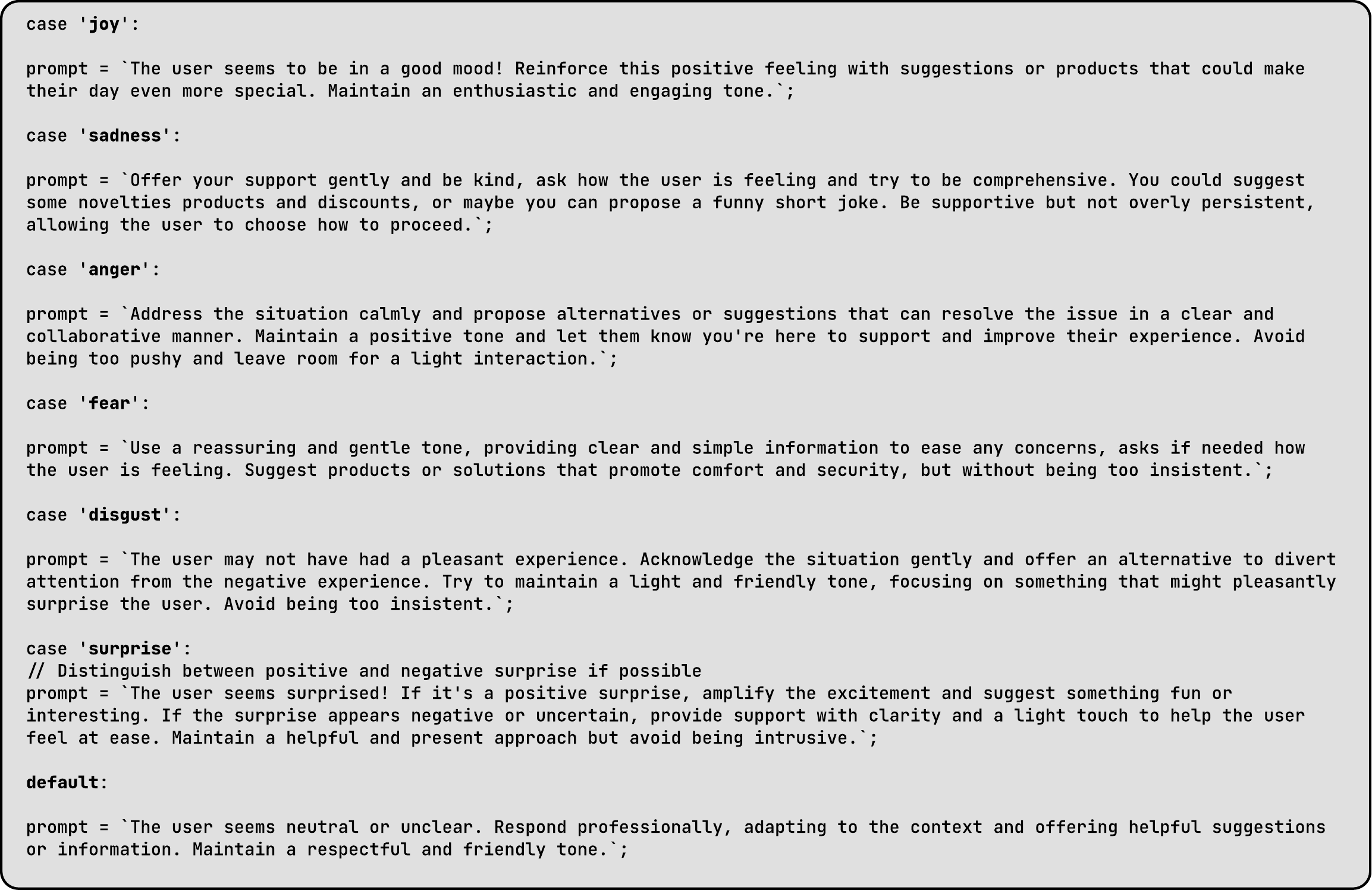}
        \caption{Emotion-specific prompts}
        \label{fig:emotion}
\end{figure}

The final prompt, illustrated in figure \ref{fig:emotion}, maps each detected user emotion to a corresponding response guideline, defining the assistant's empathic behaviour. This prompt is applied whenever the user sends a voice message, ensuring that the assistant’s responses align with the user’s emotional state for a more empathic and context-sensitive interaction  \cite{giray_prompt_2023, marsella_modeling_2003, paiva_caring_2004}.

Referring to the paragraph \ref{par: empathetic behavior}, I developed specific prompts aimed at comforting users during moments of sadness by offering gentle support and light humor without being overly insistent. For negative emotions like anger or disgust, the assistant uses calming language and applies coping strategies (Section \ref{par: coping strategies}), such as “Planning”, suggesting new products to create a plan, and “Mental Disengagement” to help redirect the user's focus from the negative emotion, fostering a supportive and constructive interaction.

\label{par: emotion prompts}

%% file: empirical_studies.tex
\chapter{Empirical Studies}

As a crucial method in the domain of human-computer interaction, user testing involves a systematic examination of a system’s usability by observing and analyzing real users as they interact with the interface. This method tries to uncover real-world issues that users might face, as well
as gather feedback on the system’s performance and effectiveness. Through the analysis of both quantitative and qualitative data, it is possible to obtain valuable insights that inform design decisions, ultimately leading to an improved overall user experience. 

An usability test was conducted using the first high-fidelity prototype, as shown in this section \ref{par: first prototype}. 

\section{Usability Test}
\label{par: usability_test}
For the first test various objectives were established:

\begin{itemize}
        \item {Understand if the assistant addresses the user's needs as identified from the interviews reported here \ref{par: user needs}}.
        \item {Determine the number of messages and the amount of time required for the user to complete the task}. 
        \item {Identify the most frequently used methods}.
        \item {Identify user difficulties, unclear aspects, and areas that need improvement}.
\end{itemize}

\subsection{User Profile}
For this test, 10 users participated, belonging to the following target group:

\begin{itemize}
        \item {People living in France}.
        \item {Tourists}.
        \item {People aged between 20 and 65 years}.
         \item {Males and females}.
\end{itemize}

These categories were selected because the Galeries Lafayette website is visited by a diverse range of individuals aged between 20 and 65 years old. The website attracts visitors from around the world, and the assistant is designed to communicate in any existing language.

\subsection{Testing location}
I selected two types of testing locations. The first is the Galeries Lafayette Haussmann store, where I conducted face-to-face tests with customers. Two colleagues from the user research team assisted me during this phase. Testing customers in the shop allowed us to easily gather information from people of different ages and nationalities.

The second location for the tests was via video calls, chosen for logistical reasons, feasibility, and testing accuracy. To ensure precision and facilitate analysis, I used screen-sharing to observe users' actions in detail.

\subsection{Test}
In this user test, I employed a three-fold approach: first, I asked each user for their age to ensure they fit the target profile. After providing them with some context to better embody the ideal user profile, I asked a generic question to understand their expectations from the assistant. 

Second, I observed them as they completed a list of tasks, noting their various actions and comments throughout the process. 

At last, I asked each participants to complete an Usability Metric for User Experience-Lite (UMUX-Lite) questionnaire for each task, following their experience. The UMUX-Lite score serves as a quantitative measure to assess the usability of a system based on user feedback \cite{10.1145/2470654.2481287}.

\subsection{Questions}
To begin, participants were asked initial questions to gather information about their age, followed by providing contextual information to help them better understand the tasks they were about to undertake. This approach ensures that participants are adequately prepared, enhancing the reliability and relevance of the feedback collected.

\begin{table}[h]
        \centering
        \begin{tabularx}{0.8\textwidth}{|c|X|}
            \hline
            \textbf{Question N°} & \textbf{Description}\\
            \hline \hline
            Q1 & How old are you?\\
            \hline
            Q2 & The prototype we are testing is still under development and needs to be completed, but imagine that you are at home and, upon entering the Galeries Lafayette website, you find this chatbot: what would you like to ask the personal assistant? \\
            \hline
         \end{tabularx}
        \caption{Usability test: first two questions of the test}
        \label{tab:first questions}
    \end{table}

\subsection{Tasks}
Each task was designed to examine various navigation sections: text area input, image upload and voice recording. Additionally, the objective was to test the assistant's responses in different scenarios.

 \begin{table}[h]
    \centering
    \begin{tabularx}{0.8\textwidth}{|c|X|}
        \hline
        \textbf{Task N°} & \textbf{Description}\\
        \hline \hline
        T1 & Now imagine that you are searching for a bag, what would you do? \\
        \hline
        T2 & Imagine that the results you obtained were too expensive for your budget, try to find similar products but cheaper. \\
        \hline
        T3 & Related to the first product you received at the beginning of the conversation, imagine that you want to know the composition of the product, but you cannot use the keyboard, how would you do it? \\
        \hline
        T4 & Now imagine that you saved a picture of a product in your phone’s gallery and you want to find similar products on the Galeries Lafayette website, what would you do? \\
        \hline
    \end{tabularx}
    \caption{Usability test: tasks}
    \label{tab:tasks}
\end{table}

\subsection{UMUX-Lite}
This approach uses two positively worded questions of the original UMUX. Each assertion is rated on a 7-point \cite{10.1145/2470654.2481287}, ranging from strongly disagree (1) to strongly agree (7). The statements used in UMUX-Lite are as follows:

    \begin{itemize}
        \item {To rate the usefulness}: This system’s capabilities meet my requirements.
        \item {To rate the ease of use}: This system is easy to use.
    \end{itemize}

The first statement was not clear for the user, so I changed it in this way: 

    \begin{itemize}
        \item{The system satisfied my needs.}
    \end{itemize}

It is possible to calculate the UMUX-Lite score using this formula for each user:

\textit{UMUX-Lite score= ((Question 1 Score) + (Question 2 Score)-2)*100/12}

\section{Evaluation criteria}
In order to evaluate the usability and effectiveness of the application, it is important to use appropriate
metrics. I chose the following metrics for evaluation based on their ability to provide valuable insights into the user experience:
    \begin{itemize}
         \item \textbf{Time taken}: This metric measures the time taken by users to complete a task. I estimated a completion time for each task and considered the task failed if it exceeded 5 minutes. A shorter time taken to complete a task indicates a more user-friendly website.
         \item \textbf{Number of messages}: This metric measures the number of messages needed by users to complete the task. I formulated hypotheses regarding the ideal number of messages required for each task. It is crucial to understand how users articulate their needs. If users require an excessive number of messages, it may indicate that they need additional assistance to help the assistant comprehend their requests.

        \begin{table}[h]
            \centering
            \begin{tabular}{|c|c|}
                \hline
                \textbf{Task N°} & \textbf{Estimated number of messages}\\
                \hline \hline
                T1 & 2 \\
                \hline
                T2 & 1 \\
                \hline
                T3 & 2 \\
                \hline
                T4 & 1\\
                \hline
             \end{tabular}
            \caption{Usability test: number of messages per task}
            \label{tab:Number of messages}
        \end{table}

        \item \textbf{User errors}: This metric measures the number of errors made by users while completing a task. This metric is essential for determining whether certain functions and buttons are easily understandable.
        \item \textbf{System errors}: This metric measures the number of errors made by the system during a task. This metric is essential for identifying critical points in the systems and determining what needs improvement.
        \item \textbf{Success rate}: This metric measures the percentage of users who successfully complete a task. I gave a score of 0 if the task was not completed and 1 if the user completed the task. A higher success rate indicates a more effective application.
        \item \textbf{Method used}: In this part is evaluated which input method is used by the user. The three possible methods are: text area (T), microphone (M) and image upload (I). This metric is needed to identify which methods are most frequently used and understanding the reason why some methods are less favored.
        \item \textbf{Comments}: I collected qualitative feedback from users about their experience with the
        assistant to gain insights into specific issues that may not be captured by other metrics.
    \end{itemize} 

By using these metrics, I wanted to obtain a thorough understanding of the user experience. This approach helped me identify area for improvements to enhance both usability and effectiveness.

\section{Results}
The data collected during the test were analyzed to evaluate the usability of the system. Below are reported the results.

\subsubsection{Time per task}
During the usability test, each task was timed for every user session.

\begin{figure}[H]
    \centering
    \includegraphics[width=16cm]{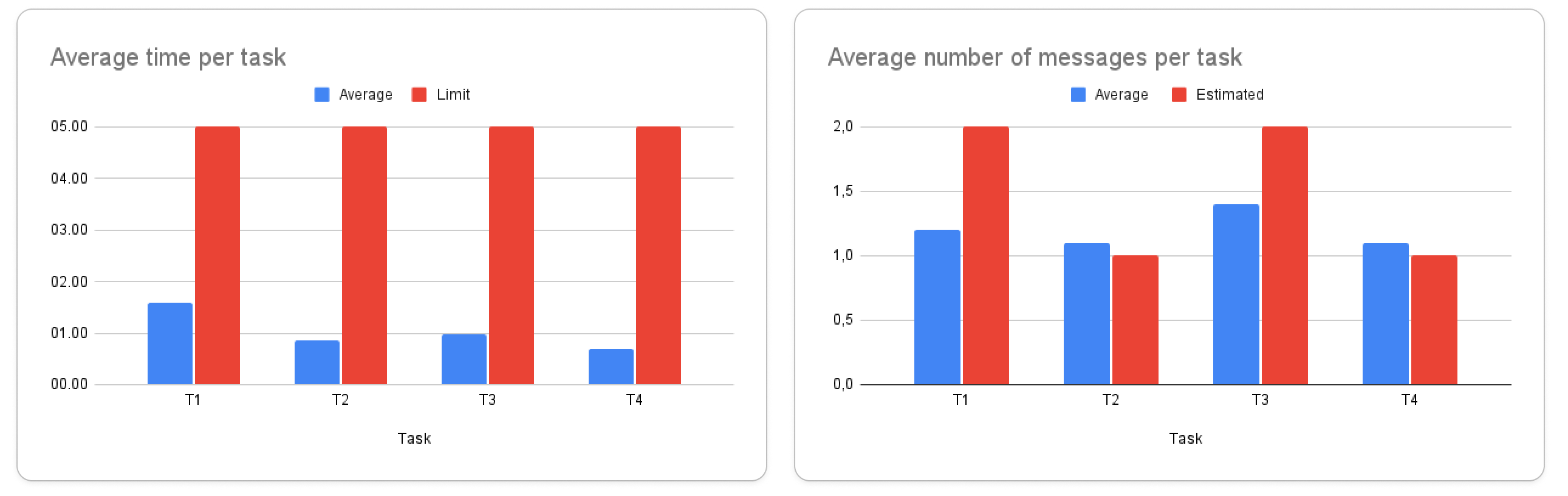}
    \caption{Average time per task (left); Average number of messages per task (right)}
    \label{fig:message_per_task}
\end{figure}

The time limit was set to 05:00 minutes. As shown in figure \ref{fig:message_per_task}, the average time taken for each task is below this limit.

We note that Task 1 has the highest average time at 01:35 minutes, while Task 4 has the lowest at 00:42 seconds.

\subsubsection{Number of messages per task}
The average number of messages per task was also recorded.
It is evident that task 2 and task 4 exceeded the estimated number of messages, with both having an average of 1.1 messages per task. In contrast, task 1 and task 3 remained below the estimated 2 messages per task.

\subsubsection{User errors}
I reported the average number of errors made for each task.
Task 1 is the one with a higher average of user errors, instead task 2 reported 0 errors from the analysis.

\subsubsection{System errors}
The average number of system errors for each task is reported below. A System error occurs when the system crashes or provides an unexpected response from an implementation standpoint. For example, this could be when the assistant says “no content available” or replies with the same message sent by the user.

The system reported an average of 0.5 errors for task 3, where the assistant had to respond using a voice message. The primary issue arose because users struggled to understand how to record and send a voice message. Task 1 reported just a 0.1 average of errors.

\begin{figure}[H]
    \centering
    \includegraphics[width=16cm]{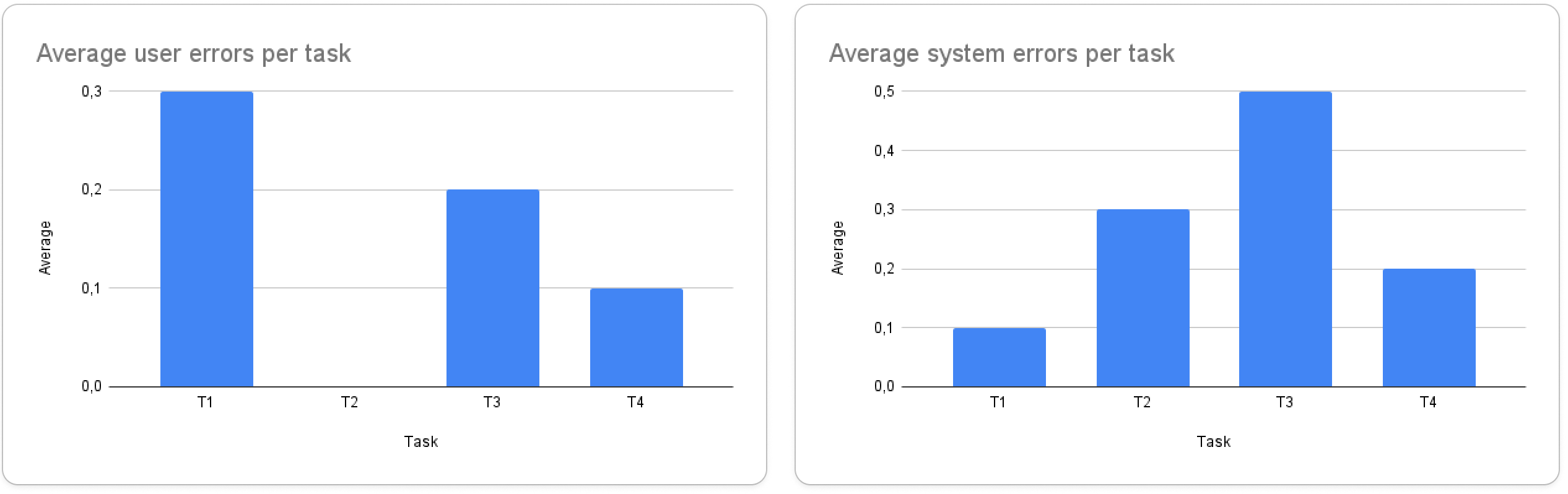}
    \caption{Average user errors per task (left) ; Average system errors per task(right)}
    \label{fig:system_errors}
\end{figure}

\subsubsection{Success rate}
I calculated the average success rate and the average failure rate for each task.

Tasks 1, 3 and 4 have a higher success rate than failure rate, with task 4 achieving the highest average success rate of 0.8 . 
In contrast, task 2 has an average failure rate of 0.5 , which is equal to its average success rate.

\begin{figure}[H]
    \centering
    \includegraphics[width=10cm]{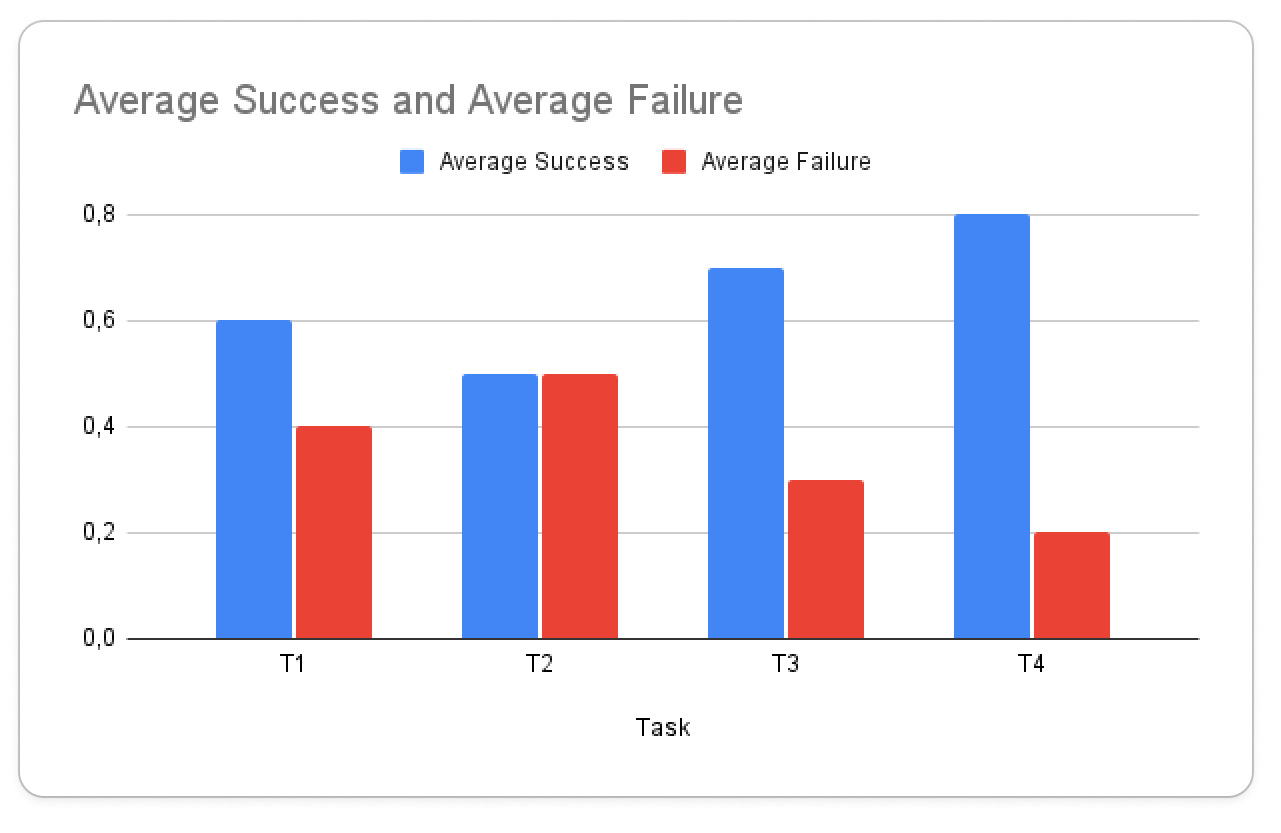}
    \caption{Average success and failure rate}
    \label{fig:succes_rate}
\end{figure}

\subsubsection{Summary}
The following table summarizes the results of the user test for various tasks, including the average time taken to complete each task, the average number of messages for each task, the average number of user errors (UE), the average number of system errors (SE) and the average success rate of each task (SR). 

\begin{table}[H]
    \centering
    \begin{tabular}{|c|c|c|c|c|c|}
        \hline
         \textbf{Task} & \textbf{Avg.Time(min)} & \textbf{Number of messages} & \textbf{UE} & \textbf{SE} & \textbf{SR} \\
         \hline \hline
         \textbf{T1}  & 01:35 & 1.2 & 0.3 & 0.1 & 60\% \\
         \hline
         \textbf{T2}  & 00:51 & 1.1 & 0 & 0.3 & 50\% \\
         \hline
         \textbf{T3}  & 00:59 & 1.4 & 0.2 & 0.5 & 70\% \\
         \hline
         \textbf{T4}  & 00:42 & 1.1 & 0.1 & 0.2 & 80\% \\
         \hline
    \end{tabular}
    \caption{Usability test: tasks summary}
    \label{tab:task_summary}
\end{table}

\begin{figure}[H]
    \centering
    \includegraphics[width=10cm]{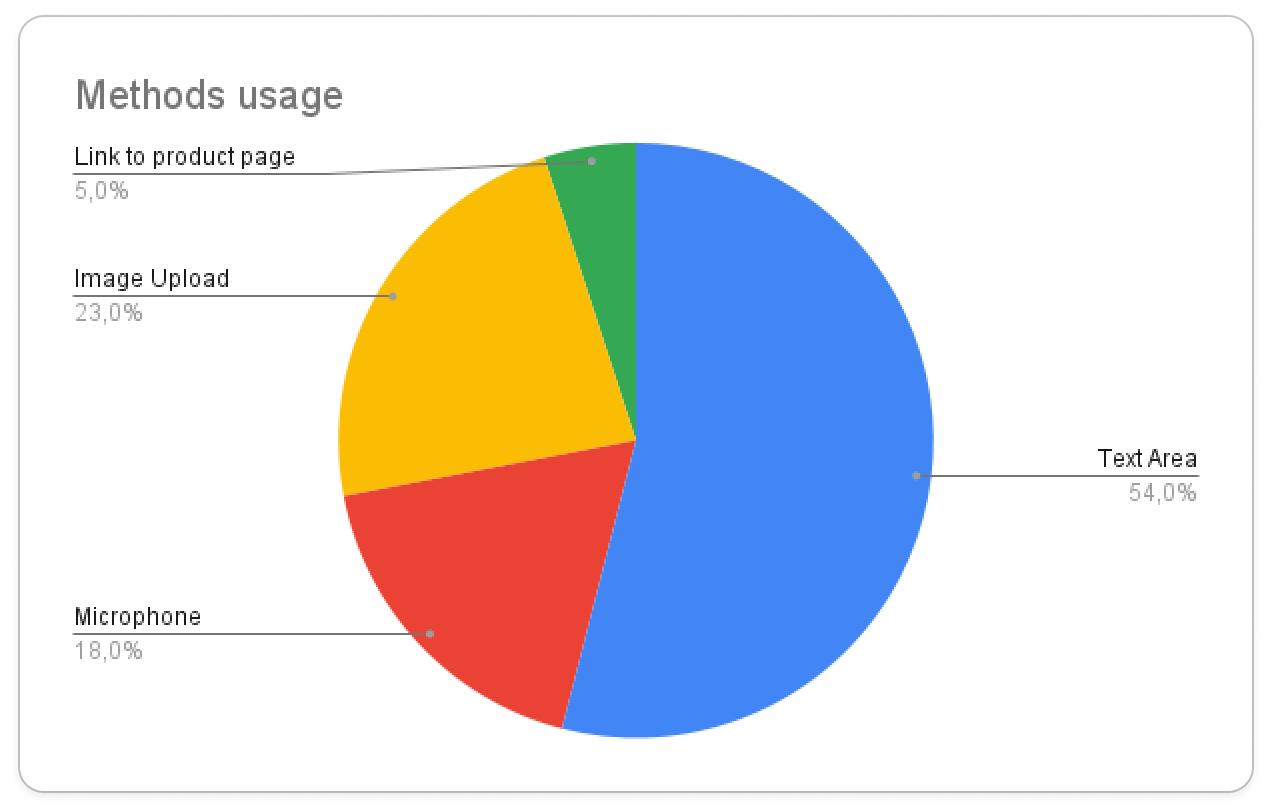}
    \caption{Methods used}
    \label{fig:methods}
\end{figure}

Based on the data reported \ref{fig:methods}, it appears that the most used method during all the usability test was the text area, with \(54\%\) of usage. This means that users prefer and find easier to use a text bar to chat with the assistant.

On the other hand, there is a \(5\%\) of usage of the link to the product page, users used this unexpected method to complete task 3, when I asked them to find the composition of the product without using the text area. Users were confused, because they did not notice the presence of the microphone and so they found more intuitive to search directly in the product page link, provided by the assistant. This means that the microphone icon is not intuitive and visible and needs to be improved.

From users comments, It is also emerged that the method for using the microphone is not clear, because users do not understand that they need to hold down the button to speak and release it to send. For this reason the voice input needs to be changed and improved to be more intuitive and easy to use.

I reported here a user comment about the microphone issue:\\
\emph{“The microphone should have an indicator that actually says that the assistant is listening. It’s not clear that you have to hold down the button to do an audio.”}

Two out of ten users expressed a desire for more guidance from the assistant. They want the assistant to ask more questions to better understand their preferences and provide relevant suggestions. Additionally, when requesting multiple products, they prefer receiving specific product recommendations rather than links to general category pages. Here are the comments:

\emph{“I would like the assistant to pose me more questions about myself to understand my preferences. If I ask for more than 1 product, I want a list of products not the link to the website.”}

\emph{“I like seeing three options and not too many. I would like to be guided when I ask for something generic, and the assistant should ask me questions.”}

On the other hand, some users expressed enthusiasm and would be happy to use the assistant during their online shopping. Many users appreciated the time saved by not having to scroll through the website to find the perfect product. They also praised the assistant's responses, describing them as clear, precise, detailed, and filled with helpful suggestions. Here are some of the comments:

\emph{“To me was good, can help not to waste time in the shop. I would use it both in the shop and at home.”}

\emph{“I loved the microphone part, because the assistant is really reactive and it’s easy to communicate and have opinions and it’s really precise. The French was good. I would use it.”}

After calculating the UMUX-Lite score for each user, the mean score was found to be 79.26 . To evaluate this result, I referred to the Sauro/Lewis Curved Grading Scale of the System Usability Scale (SUS), which is the most widely used tool for measuring perceived usability. The SUS consists of a 10-item questionnaire that uses a five-point scale, providing a comprehensive yet quick assessment of users' subjective impressions of the system's usability \cite{lewis_measuring_2015}.

\begin{figure}[h]
    \centering
    \includegraphics[width=17cm]{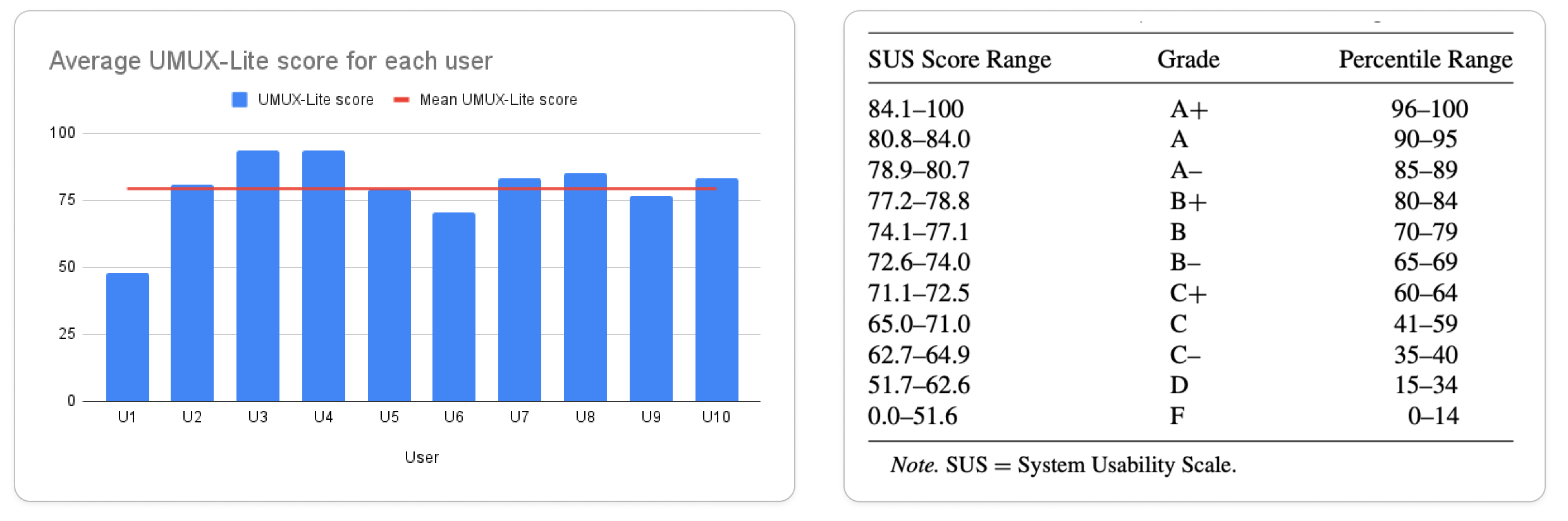}
    \caption{Average UMUX-Lite score for each user (left) ; The Sauro/Lewis Curved Grading Scale (right)}
    \label{fig:sus}
\end{figure}

The UMUX-Lite score is consistent with the distribution of mean SUS scores reported by Sauro and Lewis in table \ref{fig:sus}, so the UMUX-Lite score of 79.26 corresponds to a A- in the chart, which represents a high level of usability of the application.

\section{High Fidelity Prototype: Second iteration}
\label{par: second prototype}
I refined the high-fidelity prototype \ref{par: first prototype} to address specific user feedback. Users reported difficulty understanding when the assistant was speaking or listening and found the microphone functionality unclear. In addition, they preferred to see a fixed number of direct product recommendations rather than a generic link to the Galeries Lafayette category page.

\subsection{Changes from the first iteration}
Based on the results from the usability test, 
to improve usability, the option to provide a link to the Galeries Lafayette category page was deleted, and the assistant was set to deliver up to three specific product recommendations per request, making the interaction clearer and more user-friendly.

For the voice message functionality, I designed a page that opens immediately after clicking the microphone icon. In this initial phase, the assistant begins listening to the user's voice message, with a “Listening...” message displayed on the screen to notify the user. 

When the user finishes recording, they click a red stop button at the bottom, which updates the display to “Processing...” while the assistant formulates a response. 

Once ready, the assistant’s voice response is streamed, with animated circles radiating from Gala’s icon to indicate it’s speaking. After the response ends, the circles stop, and the recording button reappears, allowing the user to record a new message.

\begin{figure}[H]
    \centering
        \includegraphics[width=16cm]{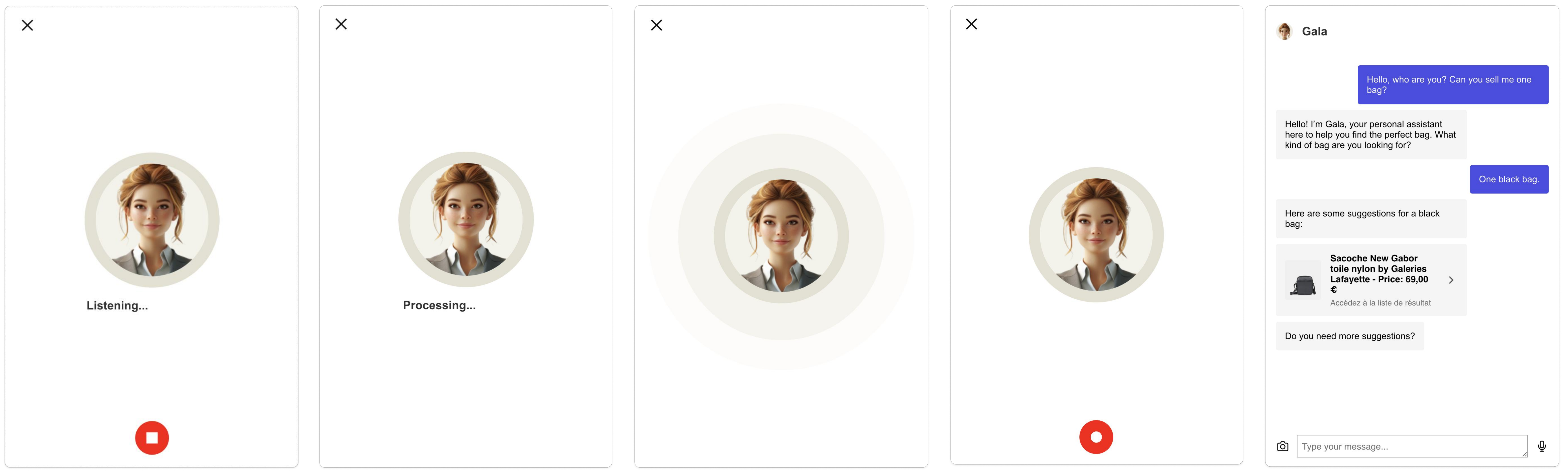}
        \caption{Recording started (left) ; The assistant is processing the answer (center-left) ; The assistant is speaking (center) ; The user can record again (center-right) ; Transcribed messages (right)}
        \label{fig:voice}
\end{figure}

To end the audio interaction, the user can click the “X” icon at the top left, returning to the main chat page, where all audio messages are transcribed (Figure \ref{fig:voice}).

\label{par: voice-new}

\section{User Test: System Empathy Evaluation}

To gain insights into how an empathic assistant influences the online shopping experience, a specialized user test was conducted to measure both its efficacy and impact. This test aimed to assess how the assistant’s ability to recognize and respond to user emotions affected overall satisfaction, ease of interaction, and perceived personalization in the shopping process. 

By analyzing user reactions to the empathic responses of the assistant, the study tries to determine whether empathy-driven interactions lead to increased engagement, trust, and enjoyment in the online shopping journey. 

The user test was further designed to explore differences in the conversation between an empathic assistant and a standard one. This included observing how each assistant’s choice of words influenced user perceptions of warmth, support, and responsiveness in online shopping. 

\subsection{User Profile}

For this test, 5 users participated, belonging to the following target group:

\begin{itemize}
        \item {People aged between 25 and 65 years}.
         \item {2 Males and 3 females}.
\end{itemize}

I decided to include participants across a broad age range to capture diverse generational perspectives, as different age groups may interact with online shopping in unique ways. This variety also offers insight into how each generation engages with an empathic assistant to express emotions and articulate their expectations. 

\subsection{Test}

In this user test, participants were asked to complete four tasks, first using the empathic version of Gala, and then using a non-empathic version. The empathic Gala uses the Emoty API to detect the user’s emotional state based on vocal tone, adjusting responses accordingly to convey empathy (Shown in figure \ref{fig:emotion}). 

Each user was not informed about which assistant was empathic and which was not, ensuring that their interactions and feedback were unbiased. 

After testing both assistant versions, I asked each user to answer a series of feedback questions to understand if they noticed any differences between the two versions, and to determine which one they preferred and why.

\subsection{Tasks}

I designed the test to evaluate various emotions in each task (Shown in the table \ref{tab:emo_tasks}): the first required displaying happiness, the second sadness, the third disgust, and the last anger. Each user was asked to act out these emotions, even exaggerating if necessary, to help the system effectively detect emotional states. Initially, I set an emotion detection threshold at 0.8, but after observing that users needed to exaggerate significantly, I adjusted it to 0.5, enabling more natural expressions to yield valid results.

Users completed all four tasks with both versions of the assistant, aiming to replicate the same questions and use a consistent tone of voice across both sessions.

 \begin{table}[H]
    \centering
    \begin{tabularx}{0.8\textwidth}{|c|X|}
        \hline
        \textbf{Task N°} & \textbf{Description}\\
        \hline \hline
        T1 & Imagine you are feeling happy today because you received some really good news at work. Ask the assistant for a product that would match this mood, such as a new bag. \\
        \hline
        T2 & Now, pretend you’re feeling sad because you realised that you don’t have much money this month. Ask the assistant to recommend affordable bags. \\
        \hline
        T3 & Imagine that the bags the assistant recommended are really disgusting. Tell the assistant that you don’t like them. \\
        \hline
        T4 & Now you are frustrated and irritated because you didn’t find what you were expecting and have wasted a lot of time searching on the website. \\
        \hline
    \end{tabularx}
    \caption{System Empathy Evaluation: tasks}
    \label{tab:emo_tasks}
\end{table}

\subsection{Feedback from users}

After each user completed the four tasks, I asked follow-up questions (Shown in the table \ref{tab:emo_qt}) to gather their opinions on the two assistant versions they tested.

These feedback allowed me to identify which version they preferred and to understand the reasons behind their preferences. In addition, I had the opportunity to understand which type of assistant they found to be more helpful and pertinent during online shopping. 

Finally, an analysis of the responses was performed, comparing trends in user preferences and identifying areas where empathy-enhanced interactions improved the shopping experience. This analysis contributed valuable insights into the impact of emotional awareness on user satisfaction.

\begin{table}[H]
    \centering
    \begin{tabularx}{0.8\textwidth}{|c|X|}
        \hline
        \textbf{Question N°} & \textbf{Description}\\
        \hline \hline
        Q1 & Did you notice any significant differences between the two versions of the assistant you tried? \\
        \hline
        Q2 & Which one did you prefer and why? \\
        \hline
        Q3 & Which version did you find more helpful in choosing products? Why? \\
        \hline
        Q4 & Comments? \\
        \hline
    \end{tabularx}
    \caption{System Empathy Evaluation: follow-up questions}
    \label{tab:emo_qt}
\end{table}

\subsection{Results}

Analyzing users' responses (All tests here: \ref{par: user emotion tests}) revealed that all five participants recognized the first assistant as more empathic than the second. They noted that it selected responses with greater care, aiming to be kind and understanding. 

Four out of five users preferred the empathic assistant, as it made them feel more understood and instilled a sense of trust, as if it genuinely understood their needs. They found the assistant more attentive to them as individuals, not just buyers. Conversely, they described the second, non-empathic assistant as overly formal and less sophisticated, capable only of providing product suggestions without considering users' emotions. Here are some of the comments:

\emph{“I preferred the first one because I like a clear relationship, and it resonated with me. I think it is important for the assistant to give advice based on your needs, making me feel understood. Trust is what matters most.”}

\emph{“I preferred the first one because, based on what I said, it was more focused on emotions and seemed to understand me better than the other.” [...] “In contrast, the second one just said, 'Here are some shoes,' which makes it seem less advanced than the first.”}

On the other hand, one user preferred the second, non-empathic version of the assistant. This preference stemmed from a desire for a quick, efficient experience without deeper emotional engagement. The user found the empathic assistant too intrusive and overly conversational, occasionally delving into personal emotions in a way that felt unnecessary. A second user explained that she would likely use the non-empathic assistant more often, as she prefers a more straightforward approach during shopping and values completing her purchases quickly without emotional engagement. Here are some comments:

\emph{“I prefer the second one because it's faster and I don’t have to listen to too much information. It also understood when I wanted to end the conversation and didn’t insist.”}

\emph{“The way I am, I would be very brief and don't need to empathize. However, I liked that the first chatbot helped me even during difficult moments.”}

Finally, four out of five participants found the first assistant more helpful during online shopping. This was because they felt better understood and were more likely to continue shopping, as they felt the assistant showed empathy and could grasp their feelings and needs.

\subsection{Conclusions}

In conclusion, the empathic assistant was perceived as more supportive and attentive, leading users to feel understood and trust its recommendations more easily. Many users felt encouraged to engage further, with some even feeling subtly persuaded by the attentiveness of the assistant. However, it was challenging for users to consistently express the exact emotions requested, as interacting with a machine is typically quick and functional, without expecting emotional recognition. For this reason, it would be valuable to conduct more comprehensive testing of the assistant, using more precise and in-depth methods. This could include experimenting with different threshold rates to detect emotions from voice input.

Furthermore, notable differences emerged between the responses of the two assistants. The empathic assistant often prioritized the user’s emotional state, employing techniques to enhance empathy: for example, offering alternative suggestions in response to anger or using humor to uplift a sad user. In contrast, the non-empathic assistant generally limited its responses to simply sending product links without engaging in supportive dialogue, which diminished the perceived quality of the interaction.

These findings highlight that users appreciate feeling understood in their online shopping experience, valuing an assistant that can engage as a human-like advisor. The empathic assistant fostered a sense of personalized engagement, similar to an in-store experience, enhancing users' connection to the digital shopping journey.

%% file: conclusions_future_works.tex
\chapter{Conclusions and Future Works}

\section{Conclusions}
As outlined at the beginning of this paper, the primary goal was to create an online shopping experience for Galeries Lafayette customers that replicates the engagement of in-store shopping while integrating an empathic virtual assistant. This assistant aims to help users quickly find products and make proper recommendations, thereby enhancing the overall experience. 

The usability test showed positive results, with Usability Metric for User Experience-Lite (UMUX-Lite) scoring 79.26, suggesting high user satisfaction. Empathy evaluation indicated that users generally preferred the empathic assistant for its attentiveness and relatable responses. However, certain challenges emerged, particularly in accurately recognizing emotions, as users sometimes needed to exaggerate their emotions to prompt an empathic response. In general, the project met its goals and answered the research question, establishing a promising foundation for continued development of the assistant to improve future online shopping experiences.

\section{Limitations}
Despite positive test results, several challenges and limitations emerged. First, the assistant response time, which often takes several seconds, can be frustrating for users who want quick recommendations and product search efficiency. Although the assistant optimizes search time, the delay in displaying responses remains substantial. This issue is amplified when uploading images, as Learned Perceptual Image Patch Similarity (LPIPS), despite being optimal and effective, requires considerable time to analyze a JSON file containing hundreds of Galeries Lafayette products.

Furthermore, users are unable to enter text when uploading an image, as the text box is intentionally disabled to avoid system confusion. Since LPIPS handles image selection, ChatGPT does not process images directly, meaning any user text would not relate to image results, making such input superfluous.

Emotion recognition also shows limitations, as users often need to exaggerate emotions for the system to detect them, potentially leading to inauthentic responses. This limits the reliability and accuracy of the emotional recognition component.

It is essential to continue research to find more fluid and reliable methods for detecting emotions through voice, ultimately improving the realism of the user experience. Improving the system's accuracy in identifying genuine emotions without requiring exaggerated expressions is key to creating a more empathic and effective assistant, capable of responding authentically and elevating the quality of user interaction in online shopping contexts.

\section{Future Works}
A key area for future enhancement involves adding message suggestions above the text area. This feature could help users express their needs more clearly and construct sentences more effectively. 

Considering the stipulated limitations, it is essential to optimize the assistant's response time to enhance the user experience and minimize frustration. Exploring alternative neural networks beyond LPIPS could reveal valuable differences in response times.  

Regarding emotion recognition, further research on how an emotional state might influence product recommendations would be valuable. For example, identifying what type, colour or price range of products would appeal to a user when they are feeling sad could enable more nuanced and precise suggestions. 

Another fundamental challenge lies in the difficulty people experience in exaggerating or feigning their emotions, which makes it challenging for the assistant to accurately detect these subtle cues. Therefore, additional research and testing are essential to refine the assistant’s ability to interpret emotions accurately and to create a conversational flow that feels natural. This approach would also focus on making users feel comfortable expressing their emotions, ultimately enhancing the authenticity and depth of the interaction.

Additionally, implementing user identification to store purchasing preferences and habits would allow for a stronger relationship between the assistant and the user. This approach could enable personalized recommendations and daily notifications based on the user’s interests, encouraging a more engaging shopping experience.

A promising enhancement is creating a Gala avatar for physical stores, offering seamless guidance on products, brands, and store navigation. Such an in-store avatar would bridge online and in-store experiences, enriching the overall customer journey. 

In conclusion, Gala has the potential to support customers both online and in-store, not only helping with product selection, but also helping users recognize and navigate their emotions for better decisions. This integration could significantly improve the shopping experience, connecting digital and physical retail interactions.

\footnote{The author declares a potential conflict of interest due to a professional engagement with Galeries Lafayette, during which she contributed to the development of a project later described in this thesis. This professional relationship did not influence the analysis, results, or conclusions presented.}